\documentclass{aa}
\bibpunct{(}{)}{;}{a}{}{,} 

\usepackage{graphicx}
\usepackage[varg]{txfonts}
\usepackage{bigstrut}
\begin{document}

   \title{Abundance analysis for long-period variables}

   \subtitle{II. RGB and AGB stars in the globular cluster 47\,Tuc}

   \author{T. Lebzelter\inst{1}
          \and
          W. Nowotny\inst{1}
          \and
          K.H. Hinkle\inst{2}
          \and
          S. H\"ofner\inst{3}
          \and
          B. Aringer\inst{3}
          }

   \institute{University of Vienna, Department of Astrophysics,
              T\"urkenschanzstrasse 17, A-1180 Vienna, Austria\\
              \email{thomas.lebzelter@univie.ac.at}
         \and
             National Optical Astronomy Observatory,
             950 N.Cherry Avenue, Tucson, Arizona 85726, USA
          \and
             Departement for Physics and Astronomy, Division of Astronomy and Space Physics,
             Uppsala University, Box 516, 75120 Uppsala, Sweden
             }

   \date{Received ; accepted }

   \titlerunning{Abundance analysis for long period variables II.}

 
\abstract
{Asymptotic giant branch (AGB) stars play a key role in the
enrichment of galaxies with heavy elements. Due to their large
amplitude variability, the measurement of elemental abundances is
a highly challenging task that has not been solved in a satisfactory
way yet.}
{Following our previous work we use hydrostatic and dynamical model
atmospheres to simulate observed high-resolution near-infrared
spectra of 12 variable and non-variable red giants in the globular
cluster 47\,Tuc.  The 47\,Tuc red giants are independently
well-characterized in important parameters (mass, metallicity,
luminosity).  The principal aim was to compare synthetic spectra
based on the dynamical models with observational spectra of 47\,Tuc
variables.  Assuming that the abundances are unchanged on
the upper giant branch in these low-mass stars, our goal is to
estimate the impact of atmospheric dynamics on the abundance
determination.}
{To estimate abundances, we measured the equivalent widths of selected
features in observed spectra and compared the results
with predictions from a set of hydrostatic and dynamical model
atmospheres resembling 47\,Tuc AGB stars in their fundamental
parameters. Our study includes lines of $^{12}$CO, $^{13}$CO, OH,
and Na. Furthermore, we investigated the variations in line intensities
over a pulsation cycle.}
{We present new measurements of the C/O and $^{12}$C/$^{13}$C ratio
for 5 non-variable red giants in 47\,Tuc.  The equivalent widths
measured for our 7 variable stars strongly differ from the non-variable
stars and cannot be reproduced by either hydrostatic or dynamical
model atmospheres.  Nevertheless, the dynamical models fit the
observed spectra of long-period variables much better than any
hydrostatic model. For some spectral features, the variations in
the line intensities predicted by dynamical models over a pulsation
cycle give similar values as a sequence of hydrostatic models with
varying temperature and constant surface gravity.}
{Our study of the dynamical effects on abundance determination
visible in these well-characterized cluster stars prepares the ground
for the long-term goal of deriving abundances for variable AGB stars
in general.}


   \maketitle

%

\section{Introduction}

The evolution of low- and intermediate-mass stars along the red
giant branch (RGB) and the asymptotic giant branch (AGB) is accompanied
by major mixing events that change the elemental abundance pattern
derivable for the stellar atmosphere. The surface abundances then allow a view
into the nucleosynthesis processes in the stellar interior and
provide important clues to the origin of the elements in the
interstellar medium and ultimately in ourselves.  The main tool for
studying the abundance pattern is high-resolution spectroscopy. However,
deducing the element abundances from the spectra requires an
understanding of the physical conditions in the line-forming atmospheric
layers.  Two properties of red giants, in particular of those on
the AGB, complicate this procedure: very extended atmospheres and
dynamical effects due to radial pulsation.  Several authors have worked
on this question in the past, in particular, \citet{1992A&A...253..203S}, 
\citet{1996A&A...307..481B}, and \citet{2007MNRAS.378.1089M}.

In the first paper of this series \citep[][Paper\,I]{2010A&A...517A...6L},
we summarised the previous studies and presented a new investigation
of the dynamical effects on stellar abundance determinations based
on the dynamical model atmospheres by \citet{2003A&A...399..589H}.  
The observed changes in line depth over a pulsation
cycle cannot be reproduced by a simple sequence of hydrostatic
models of varying effective temperature or surface gravity, because
the pulsation leads to a clearly non-hydrostatic atmospheric structure
\citep[e.g.][]{2010A&A...514A..35N}
producing very different line depths.  The well known phenomenon
of line doubling \citep[e.g.][]{1982ApJ...252..697H} adds further
complication to this matter. The changes in line strength over a
light cycle produce a loop in an equivalent width (EW) vs. $(J-K)$
diagram; i.e., the rising and descending branch of the light curve
do not show the same line strength at the same near-infrared (NIR)
colours. This result is also found qualitatively in the observational
data of one field mira analysed in Paper\,I.

The conclusion from Paper\,I is that only a dynamical model atmosphere can
properly describe the atmospheric structure and temporal
change in highly variable stars at the tip of the giant branch.
Therefore, it is timely to ask if state-of-the-art dynamical models
can be used to derive abundances for such stars. The goal of
the current paper is to explore this question by comparing independently
well-characterized objects with dynamical model atmospheres for which the
fundamental parameters ($M$, $L$,
[Fe/H], pulsation period, and $T_{\rm eff}$)
were chosen to fit the observed targets. 

Cluster stars are an excellent testbed for such investigations.
Parameters, such as mass, metallicity, and luminosity, can be
determined on the basis of a large number of stars, 
without relying on the giants. This is a major advantage since deriving
abundances for cool giants is a problem with many degrees of freedom
so that any constraints of parameters from independent sources
reduce the uncertainties significantly. 

The study of changes in the intensity and profile of an atomic or
molecular line in cool giants is hampered by the large number of
blending spectral features. To minimize the impact of line blending,
our strategy -- previously presented in Paper\,I -- is to study
NIR lines in metal-poor objects. With plenty of atomic and
several series of molecular lines available, the NIR is
the preferred wavelength range to derive element abundances in cool
giants. In the present paper, we continue this strategy.

We decided to use NIR spectra of 47\,Tuc red giants
for our study since they combine the advantages of (i) cluster membership and (ii) reduced 
line blending. We constrain the abundances of the variable stars in advance
by deriving selected element abundances for a sample of non-variable stars in that
cluster with the help of hydrostatic model atmospheres.

\begin{table*}
\caption{Properties of the \object{47\,Tuc} stars (LPVs, non-variable RGB stars), all of which are assumed to have $M_\star$\,=\,0.6\,$M_{\odot}$ and $[$Fe/H$]$\,=\,$-0.7$\,dex 
(Paper\,I). The photometric data is described in Sect.\,\ref{47tucsample}. $K$ and $(J-K)$ are given for the corresponding pulsation phase $\phi$.
Periods are taken from \citet{2005A&A...441.1117L} (uncertain values are marked with a colon). 
The photometric variations $\Delta V$ were adopted from \citet{2005A&A...432..207L}, while the estimates for $T_{\rm eff}$ and log\,$g$ are described in Sect.\,\ref{evaluation}.
}
\label{t:sample}
\centering
\begin{tabular}{l c c c c c | c c c}
\hline\hline
Object-ID & $K$ & $(J-K)_{\rm 0}$ & Period & Pulsation & $\Delta V$ & $L_{\star}$ & $T_{\rm eff}$ & log\,$g$ \bigstrut[t]\\
 & [mag] & [mag] & [d] & phase $\phi$ & [mag] & [$L_{\sun}$] & [K] & log [cm\,s$^{-2}$] \bigstrut[b]\\
\hline
V1 & 6.20 & 1.24 & 221 & 0.90 & 4.8 & 4760 & 3410 & $-$0.2 \bigstrut[t]\\
V2 & 6.28 & 1.14 & 203 & 0.35 & 4.5 & 4470 & 3620 & $-$0.1\\
V3 & 6.26 & 1.18 & 192 & 0.18 & 3.5 & 4590 & 3540 & $-$0.1\\
V4 & 6.68 & 1.19 & 165 & 0.80 & 2.5 & 3260 & 3520 & 0.1\\
V7 & 6.96 & 1.16 & 52 & -- & 0.4 & 2450 & 3590 & 0.2\\
V11 & 6.70 & 1.15 & 160: & -- & 0.2 & 3110 & 3600 & 0.1\\
V18 & 7.46 & 1.07 & 83: & -- & 0.25 & 1550 & 3750 & 0.5 \bigstrut[b]\\
\hline
Lee 1505 & 8.53 & 0.96 & -- & --& -- & 710 & 4020 & 0.9 \bigstrut[t]\\
Lee 1510 & 8.77 & 0.91 & -- & --& -- & 610 & 4090 & 1.0\\ 
Lee 1603 & 7.95 & 1.07& -- & --& -- & 1150 & 3800 & 0.6\\
Lee 2426 & 8.50 & 0.96 & -- & --& -- & 720 & 4020 & 0.9\\
Lee 4603 & 8.75 & 0.88 & -- & -- & -- & 630 & 4170 & 1.0 \bigstrut[b]\\
\hline
\end{tabular}
\end{table*}

\section{The 47\,Tuc reference sample}\label{47tucsample}
\subsection{Observations}
47\,Tuc is one of the closest globular clusters. It has a well
populated giant branch, and the AGB stars of that cluster have
been studied in several papers of our group \citep[see][]{2005A&A...432..207L, 2006ApJ...653L.145L}.
There we reviewed the cluster's  main parameters as found in the literature.
For the present study we use a distance modulus $(m-M)$ of 13.5 mag.
Interstellar extinction and reddening is low towards this cluster,
we took a value of $E(B-V)$=0.024 and $R_V$ =3.1 as in \citet{2005A&A...432..207L}.
We use a metallicity [Fe/H]$=-$0.7. For modelling stars on the upper giant branch
of 47\,Tuc, we set the current mass to 0.6\,$M_{\sun}$ following \citet{2005A&A...441.1117L}.

From the 42 known variable red giants in 47\,Tuc \citep[see their Table
1]{2005A&A...441.1117L}, we selected a subset of seven targets for our
abundance study. The main characteristics of these long period
variables (LPVs; object identifications V followed by a number) are
listed in Table \ref{t:sample}. Table \ref{t:sample} also contains the corresponding
data for our sample of non-variable RGB stars in 47\,Tuc (object identifications
Lee followed by a number; identified originally by
\citealt{1977A&AS...27..381L}).  Two different data sets of
observational spectroscopy were used in this work.

First, we obtained single-epoch high-resolution (R$=$50\,000) NIR
spectra for the complete sample of stars using the Phoenix spectrograph
at Gemini South \citep{2003SPIE.4834..353H}. Observations were done
on three consecutive nights in December 2002 using two wavelength
settings around 1.555 and 2.341\,$\mu$m (cf.
Fig.\,\ref{f:specLee}). Each setting covers a spectral range of
approximately  100\,{\AA}. These single-epoch spectra were used to
derive general abundance information in Sect.\,\ref{s:results1}.

Second, we analysed a spectroscopic time series around 1.630\,$\mu$m
(R$=$37\,000; cf. Fig.\,\ref{f:specvariability}) for a subsample
of objects (V1, V2, V3, V4, V11, and V18), obtained at Mount Stromlo
Observatory in 2002.  These spectra were previously used in a
study of the velocity variations of the 47\,Tuc LPVs 
\citep{2005A&A...432..207L}, and we refer to that publication for
further details on the observations. The Mount Stromlo observations
included all the variables from the Gemini sample but not the RGB
stars. These time series of spectra were used to gather information
on the temporal variability of EWs in
Sect.\,\ref{s:results2}.

The observed spectra were reduced using standard techniques for
NIR spectra. Wavelength calibration was done with spectra
of K-type standard stars.  The wavelength regions were selected to
include only a negligible ($H$-band) or low ($K$-band) contamination
by telluric lines which were removed. The spectra were corrected
for radial velocity shifts (stellar motion, atmospheric kinematics) 
so that the line centres are at the
laboratory wavelengths. For line identification we used the
NIR spectral atlas of Arcturus provided by
\citet{1995iaas.book.....H}. Instrumental distortion of the
(pseudo)continuum was removed.  For the Mount Stromlo data we had
a few cases where two spectra of the same star were obtained on
consecutive nights. These spectra were not combined but analysed
separately and used for constraining the measurement uncertainty.

\subsection{Stellar parameters}
The membership of our stellar sample in a globular cluster makes 
determining some stellar parameters (e.g.~$L$, $M$, [Fe/H]) simpler and more reliable
than for field stars. A notable exception, however, is the effective temperature
that we estimate from NIR photometry.
We did not obtain simultaneous NIR photometry. For the
Lee stars we used $(J-K)$ values from \citet{1981ApJ...246..842F}.
For the small-amplitude variables V7, V11, and V18, we used the
values given in \citet{2005A&A...441.1117L}. For the large amplitude
variables, V1, V2, V3, and V4, which show a significant brightness
amplitude also in the NIR, we first derived the pulsation
phase at the time of the observation using light curve data presented
in \citet{2005A&A...441.1117L}. Based on the photometric variability
range in the NIR of our sample stars, compiled from various
individual measurements and short time series available in the
literature \citep{1973MNRAS.163..245G,1981ApJ...246..842F,
1983ApJ...272..167F, 1985MNRAS.212..783M, 1988ApJ...324..823F,
2005A&A...432..207L}, we estimated infrared colours for the calculated
pulsation phase at the time of the Gemini observations. These are
the values given in Table \ref{t:sample}.  For the large-amplitude
variables, we were able to determine the colour change with good
coverage of the whole light cycle \citep[e.g.][]{1985MNRAS.212..783M}.
However, even in those cases the total range in $(J-K)$ for any 
star does not exceed 0$\fm$15, so that a large error in our
$(J-K)$ values can be excluded. Phasing of the Mount Stromlo
measurements was done using the parallel light curves given in
\citet{2005A&A...432..207L} and \citet{2005A&A...441.1117L}.  All
photometry was transformed to the photometric system of
\citet{1988PASP..100.1134B} that was also used for calculating our
synthetic photometry.

For relating NIR colour and $T_{\rm eff}$, we used hydrostatic
models as outlined in detail in Sect.\,\ref{evaluation}. 
With the distance and turn-off mass
of 47\,Tuc, we can estimate the typical $L$, $T_{\rm eff}$, and
log\,$g$ values for our cluster stars. Non-variable stars are found
in the temperature range between 3800 and 4200\,K, and their surface
gravity log\,$g$ is between 0.6 and 1.0. For the variable stars we
get temperatures between 3500 and 3750\,K and log\,$g$ values between
$-$0.2 and 0.5. However, these estimates assume a hydrostatic
structure and therefore have to be taken with some caution. The
calculated stellar parameters for each star are given in the final
three columns of Table \ref{t:sample}. Our temperatures are in good
agreement with the values derived by \citet{2005A&A...441.1117L},
but are somewhat higher than the earlier values given in
\citet{1996MNRAS.278...11F}. Also listed in Table \ref{t:sample} are estimates
for the luminosity $L_{\star}$ at the phase of observation given there.

Mass loss is a critical question when comparing models of AGB stars
with observations. However, measurement of mass loss rates remains
a very challenging task.  Several studies have dealt with the mass
loss from the 47 Tuc AGB variables. For our sample, the
occurrence of dusty mass loss in V1 to V4 and in V18 is confirmed
both by an infrared excess
\citep[e.g.][]{2001A&A...372...85R,2011ApJS..193...23M} and by
mid-infrared spectroscopy \citep{2006ApJ...653L.145L}.  V7 seems
to be dust free \citep{2006ApJ...653L.145L}. Estimates of mass loss
rates found in the literature cluster around 10$^{-6}$ to
10$^{-5}\,M_{\sun}$yr$^{-1}$ for the stars on the upper AGB  of
this cluster
\citep{1988ApJ...324..823F,1997MNRAS.292..753O,2007A&A...476.1261M}.

\begin{figure}
 \resizebox{\hsize}{!}{\includegraphics[clip]{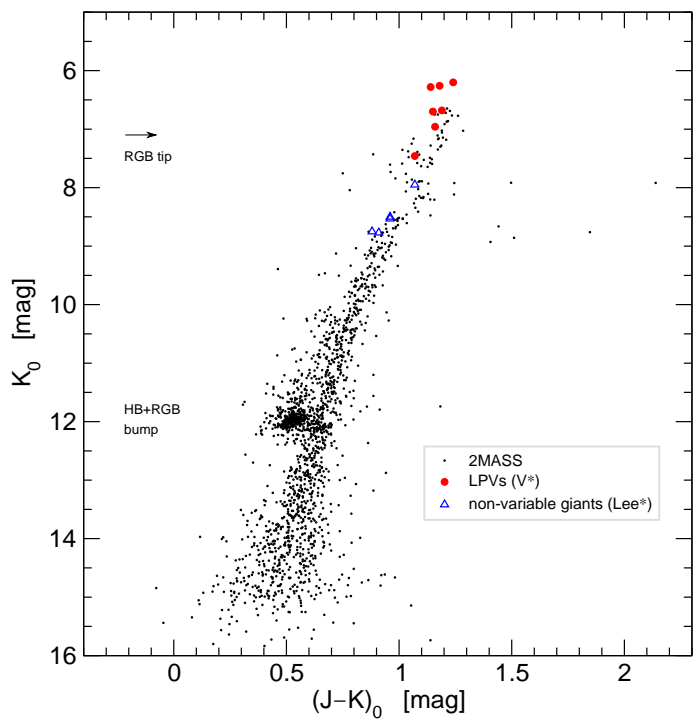}}
 \caption{Colour-magnitude diagram of 47 Tuc based on 2MASS data. The location of our sample stars are shown, divided into variable stars (red
 filled circles) and non-variable stars (blue open triangles). The photometry has been transferred to the Bessell system.}
 \label{f:cmd47}
\end{figure}

In Fig.\,\ref{f:cmd47} we show the location of our sample stars
in a colour-magnitude diagram (CMD) of the cluster.  The CMD was compiled
from 2MASS
data within 4 arc minutes of the approximate cluster centre. The 2MASS
data were dereddened and transferred to the Bessell system. The
brightness of the RGB tip is marked according to
\citet{2005A&A...441.1117L}, the location of the horizontal branch
and the RGB bump are consistent with the findings of
\citet{2007A&A...476..243S}.  All our variables are located on the
uppermost part of the giant branch, and with the exception of V18, they are all
above the RGB tip, i.e. they are AGB stars. The non-variable stars
in our sample are either on the early AGB or on the RGB.

%
\section{Methods}

We use the same approach as in Paper\,I to characterize both the
observed spectra of 47\,Tuc stars and the synthetic ones computed
for hydrostatic and dynamical atmospheric structures, namely the
measurement of EWs, of selected atomic and molecular
lines, and of molecular band heads. The spectroscopic features were
selected to determine various 
aspects of the abundance pattern in the 47\,Tuc stars, namely (i)
the C/O ratio, (ii) the isotopic ratio of carbon $^{12}$C/$^{13}$C,
(iii) the ratio of two transitions of the same species with
different excitation of the ground level (radial atmospheric
structure), and (iv) the abundance of an atomic species (Na). 
For the study of (v) the feature variability over a
pulsation cycle, we selected the CO 4-1 P27 line, which is the least
blended line within the spectral range covered by the Mount Stromlo
observations.

Standard methods within IRAF were used to measure EWs. For
selecting the appropriate measurement range around each line
centre, we considered the maximum feature width observed in any of
the analysed hydrostatic spectra (see also Sect.\,\ref{uncertain}).
In the case of the measured band heads, the short wavelength limit
was set to the cut-off wavelength while the long-wards wavelength
limit was set at an arbitrary position within the band head. All
spectra were analysed with the same set of measurement limits for
the studied spectral features as listed in Table \ref{t:features}.
Since we had shifted the spectra so that the line
cores always fall on the same position in wavelength (averaged over 
the observed spectral range), the measurement range could be kept 
constant and independent of pulsation phase.

\begin{table}
\caption{List of the features studied. Columns 3 and 4 give the borders of the spectral ranges applied to
determine equivalent widths (cf. Paper \,I).}
\label{t:features}
\centering
\begin{tabular}{l c c c c}
\hline\hline
Species & Transition & $\lambda_{\rm start}$ & $\lambda_{\rm end}$ & Note \bigstrut[t] \bigstrut[b]\\
\hline
OH & blend & 15568.6 {\AA} & 15572.1 {\AA} & 1 \bigstrut[t]\\
OH & 2-0 P$_{\rm 1e}$11.5 & 15572.1 {\AA} & 15574.8 {\AA} & 1\\
OH & 2-0 P$_{\rm 1f}$11.5 & 15574.9 {\AA} & 15578.6 {\AA} & 1 \\
$^{12}$CO & 3-0 band head & 15581.1 {\AA} & 15585.7 {\AA} & 1, 2\\
\hline
$^{12}$CO & 4-1 P27 & 16307.1 {\AA} & 16309.1 {\AA} & 5\bigstrut[t]\\
\hline
Na & & 23383.6 {\AA} & 23387.4 {\AA} & 4\bigstrut[t]\\
$^{12}$CO & 3-1 R18 & 23438.5 {\AA} & 23443.1 {\AA} & 3\\
$^{12}$CO & 3-1 R82 & 23443.1 {\AA} & 23445.6 {\AA} & 3\\
$^{13}$CO & 2-0 band head & 23447.1 {\AA} & 23452.3 {\AA} & 2 \bigstrut[b]\\
\hline
\end{tabular}
\tablefoot{The last column denotes the purpose of the individual features 
using the following designations: 1 $\rightarrow$ C/O ratio; 2 $\rightarrow$ $^{12}$C/$^{13}$C;
3 $\rightarrow$ atmospheric structure; 4 $\rightarrow$ [Na/Fe]; 5 $\rightarrow$ temporal variation.}
\end{table}

\subsection{Model atmospheres and spectral synthesis} \label{models}

Two different types of atmospheric models were used to calculate
synthetic spectra and photometry in this work, the parameters of
which were chosen to resemble the various 47\,Tuc objects introduced
in Sect.\,\ref{47tucsample} and Table\,\ref{t:sample}.
First, we produced a grid of hydrostatic, dust-free COMARCS model
atmospheres (O-rich). Being representative of the non-variable RGB
stars in 47\,Tuc, these models also serve as the hydrostatic reference
grid for the subsequent illustration of dynamic effects. The models
were computed for the parameter range specified in
Table\,\ref{t:comarcsparameters} as described in detail in
\citet{2009A&A...503..913A}.

Second, we used a small grid of O-rich dynamic model atmospheres
to investigate the effects of pulsation and mass loss on the resulting
spectra.  The aim was to produce models resembling the LPVs in 47
Tuc. The models combine time-dependent dynamics with frequency-dependent
radiative transfer, allowing us to take pulsation-induced shock
waves and opacities from atoms, molecules, and dust grains into
account. Stellar pulsation is simulated by varying the gas velocity
and the luminosity at the inner boundary that is located below the
stellar photosphere. The basic physical equations, assumptions, and
methods used for computing these models are described in
\citet{2003A&A...399..589H}, but some adjustments were made compared
to these earlier models to accommodate the properties of the present
stellar sample, in particular regarding the driving mechanism of
the stellar wind.

As discussed in Sect.~\ref{47tucsample}, observations indicate both
ongoing mass loss and the presence of dust in the outflows of the
47\,Tuc LPVs.  However, it is still a matter of debate as to what drives
the winds of O-rich AGB stars, in particular at low metallicity and
relatively high effective temperature, as in this sample. Therefore,
we used a parameterized description of the driving force of the
wind, suitable for mimicking radiation pressure on different dust
species or other wind mechanisms. More specifically, the current
approach is based on a parameterized treatment of dust formation
and grain opacities as described in \citet{2012A&A...546A..76B}.
Given the uncertainties regarding the driving mechanism and relevant
dust species, we computed models with two types of forces: (i)
radiation pressure due to true absorption by dust grains (which
also has significant effects on the resulting spectral energy
distribution), and (ii) an acceleration due to pure scattering on
dust grains (or another force, with a similar radial profile), not
causing significant circumstellar reddening. In terms of the formulae
given in \citet[][their Sect.~3.2]{2012A&A...546A..76B}, these two
cases correspond to $f_{\rm abs}$\,=\,1 and $f_{\rm abs}$\,=\,0,
respectively, where $f_{\rm abs}$ defines the fraction of the
dust opacity that is to be considered as true dust absorption. For
the other parameters we assume $\kappa_{\rm 0}$\,=\,3\,cm$^2$/g
(scaling factor of the overall absorption), $p$\,=\,0 (i.e.~no
dependence on wavelength) and a condensation temperature $T_{\rm
c}$\,=\,1500\,K.

Extending the set of two models presented in Paper I, we varied the stellar parameters of the grid models (cf. Table \ref{t:dmaparameters}) 
to represent a characteristic 47 Tuc mira, 
for instance V3 (upper row of the 2$\times$2-table), or an LPV in this system with less pronounced variations such as LW12 (lower row). In addition, 
we chose two different values for the C/O ratio, namely 0.48 (solar) and 0.25, as labelled in the main columns of Table \ref{t:dmaparameters}. 
For each combination of these fundamental parameters we computed dynamical models with different pulsation properties, parameterized by the period, 
the piston velocity amplitude $\Delta u_{\rm p}$, and the luminosity amplitude parameter $f_{\rm L}$ (see \citealt{2010A&A...514A..35N} for a definition, and 
\citealt{2014arXiv1404.7515E} for a discussion). This resulted in models for pulsating atmospheres (P), as well as models that, in addition, develop a wind (PM).

The radial atmospheric structures (temperature-pressure) of the
hydrostatic and dynamic
model atmospheres were used to calculate synthetic spectra under
the assumptions of chemical equilibrium, as well as conditions of
LTE following the approach of \citet{2009A&A...503..913A}. We adopted
the values for solar composition provided by \citet{1989GeCoA..53..197A},
except for C, N, and O, which were taken from \citet{1994LNP...428..196G}.
For more details on our calculation of synthetic spectra and
photometry, we refer to the extensive description in Paper\,I.

\begin{table}
\begin{center}
\caption{Parameter ranges covered by the applied grid of hydrostatic (COMARCS) model atmospheres. Similar to the approach followed in Paper\,I, we kept the other parameters constant for all the models: $M_\star$\,=\,1\,$M_{\odot}$, $[$Fe/H$]$\,=\,$-0.7$\,dex, $\xi$\,=\,2.5\,km\,s$^{-1}$.}
\begin{tabular}{lcc}
\hline
\hline
&Range:&Stepwidth: \bigstrut[t]  \bigstrut[b]\\
\hline
C/O&0.48$\quad$/$\quad$0.25& \bigstrut[t] \\
$T_{\rm eff}$ [K]&2600 $\ldots$ 4500&$\Delta$=100\\
log\,($g$ [cm/s$^{2}$])&0.0 $\ldots$ +2.5&$\Delta$=0.5  \bigstrut[b]\\
\hline
\end{tabular}
\label{t:comarcsparameters}
\end{center}
\tablefoot{
The isotopic ratio $^{12}$C/$^{13}$C was varied only for the spectral synthesis.}
\end{table}

\subsection{Evaluation of the models} \label{evaluation}

\begin{figure}
 \resizebox{\hsize}{!}{\includegraphics{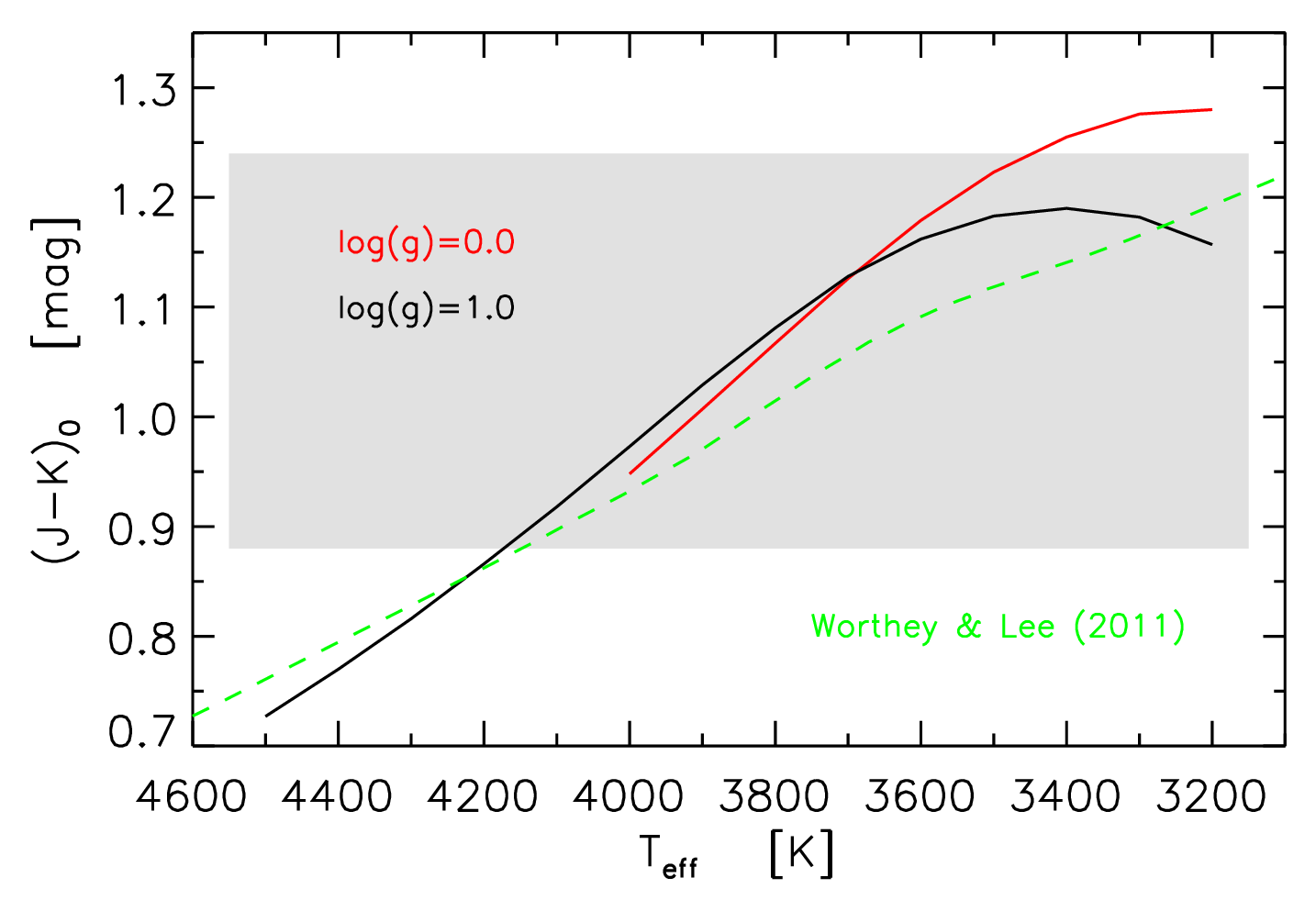}}
 \caption{Relation of near-infrared colour and effective temperature as
 derived from our hydrostatic model grid (C/O\,=\,0.25; solid lines). The green dashed
 line shows the empirical relation of \citet{2011ApJS..193....1W}. The
 shaded region marks the colour range of our observed sample of 47\,Tuc
 objects.}
 \label{f:JKTeff}
\end{figure}

In Fig. \ref{f:JKTeff} we illustrate the relation of $(J-K)$ and
$T_{\rm eff}$ as derived from our hydrostatic model atmospheres for
two different values of surface gravity. The shaded area indicates
the $(J-K)$ range of the 47\,Tuc giants studied in this paper
(see Sect.\,\ref{47tucsample}). We
note that there is a turnaround of the relations in the range 3200 to
3400\,K resulting in bluer colours for cooler stars. This result
of the modelling has been seen already in Paper\,I.  The 
figure shows a semi-empirical relation derived by
\citet{2011ApJS..193....1W}, which does not show this behaviour.
We suspect that this difference is due to the lack of dust in our
hydrostatic models.

\begin{sidewaystable*}
\caption{Characteristics of the nine dynamic model atmospheres used for the spectral synthesis listed in a 2$\times$2 table for two different sets of stellar parameters (given in the leftmost 
column) and two different C/O~ratios (given in the top row). The following parameters are coded in the names of the models (e.g. L4C048AP): luminosity (L4), C/O~ratio (0.48), the combination of 
the piston-velocity amplitude $\Delta u_{\rm p}$ and the luminosity-amplitude parameter $f_{\rm L}$ 
(A), as well as a simplified mass-loss information (P\,=\,only pulsating, PM\,=\,pulsating and mass-losing).}
\label{t:dmaparameters}
\centering
\begin{tabular}{l||ll|cc||ll|ccc}
&\multicolumn{4}{|c||}{C/O\,=\,0.48}&\multicolumn{5}{|c}{C/O\,=\,0.25} 
\bigstrut[b]\bigstrut[t] \\
\hline
\hline
$L_\star$\,=\,4000\,$L_{\odot}$ & Model: & & L4C048AP & L4C048CPM & Model: & & \textbf{L4C025AP} & \textbf{L4C025CPM1} & L4C025CPM2 \bigstrut[t]  \bigstrut[b]\\
\cline{2-10}
$M_\star$\,=\,0.6\,$M_{\odot}$ & $\Delta u_{\rm p}$ & [km\,s$^{-1}$] & 2 & 4 & $\Delta u_{\rm p}$ & [km\,s$^{-1}$] & 2 & 4 & 4  \bigstrut[t]  \bigstrut[b] \\
$T_\star$\,=\,3500\,K & $f_{\rm L}$ & & 4 & 3 & $f_{\rm L}$ & & 4 & 3 & 3  \bigstrut[t]  \bigstrut[b]\\
$[$Fe/H$]$\,=\,$-0.7$\,dex & $\Delta m_{\rm bol}$ & [mag] & 0.86 & 1.35 & $\Delta m_{\rm bol}$ & [mag] & 0.86 & 1.35 & 1.35  \bigstrut[t]  \bigstrut[b]\\
log\,$g$\,=\,--0.26 & $\kappa_{\rm 0}$ & [cm$^{2}$\,g$^{-1}$] & 0 & 3 &  $\kappa_{\rm 0}$ & [cm$^{2}$\,g$^{-1}$] & 0 & 3 & 3  \bigstrut[t]  \bigstrut[b]\\
 & $f_{\rm abs}$ & & -- & 1 & $f_{\rm abs}$ & & -- & 1 & 0   \bigstrut[t]  \bigstrut[b]\\
$P$\,=\,200\,d & $\langle\dot M\rangle$ & [$M_{\odot}\,$yr$^{-1}$] & -- & 2.0\,$\times$\,10$^{-7}$ &  $\langle\dot M\rangle$ & [$M_{\odot}\,$yr$^{-1}$] & -- & 1.5\,$\times$\,10$^{-7}$ & 5.0\,$\times$\,10$^{-7}$  \bigstrut[t]  \bigstrut[b]\\
& $\langle u \rangle$ & [km\,s$^{-1}$] & -- & 13.0 & $\langle u \rangle$ & [km\,s$^{-1}$] & -- & 13.8 & 14.5  \bigstrut[t]  \bigstrut[b] \\
\hline\hline
$L_\star$\,=\,3000\,$L_{\odot}$ & Model: & & L3C048AP & L3C048BP & Model: & & \textbf{L3C025AP} & \textbf{L3C025BP}  \bigstrut[t]  \bigstrut[b]\\
\cline{2-10}
$M_\star$\,=\,0.6\,$M_{\odot}$  & $\Delta u_{\rm p}$ & [km\,s$^{-1}$] & 2 & 3 & $\Delta u_{\rm p}$ & [km\,s$^{-1}$] & 2 & 3   \bigstrut[t]  \bigstrut[b] \\
$T_\star$\,=\,3550\,K  & $f_{\rm L}$ & & 4 & 4 & $f_{\rm L}$ & & 4 & 4  \bigstrut[t]  \bigstrut[b]\\
$[$Fe/H$]$\,=\,$-0.7$\,dex & $\Delta m_{\rm bol}$ & [mag] & 0.59 & 0.91 & $\Delta m_{\rm bol}$ & [mag] & 0.59 & 0.91  \bigstrut[t]  \bigstrut[b] \\
log\,$g$\,=\,--0.11 & $\kappa_{\rm 0}$ & [cm$^{2}$\,g$^{-1}$] & 0 & 3 &  $\kappa_{\rm 0}$ & [cm$^{2}$\,g$^{-1}$] & 0 & 3  \bigstrut[t]  \bigstrut[b] \\
 & $f_{\rm abs}$ & & -- & 0 & $f_{\rm abs}$ & & -- & 0    \bigstrut[t]  \bigstrut[b]\\
$P$\,=\,120\,d  & $\langle\dot M\rangle$ & [$M_{\odot}\,$yr$^{-1}$] & -- & -- &  $\langle\dot M\rangle$ & [$M_{\odot}\,$yr$^{-1}$] & -- & --  \bigstrut[t]  \bigstrut[b]\\
 & $\langle u \rangle$ & [km\,s$^{-1}$] & -- & -- & $\langle u \rangle$ & [km\,s$^{-1}$] & -- & --  \bigstrut[t]  \bigstrut[b] \\
\hline
\end{tabular}
\tablefoot{Notation adopted from previous works \citep{2010A&A...517A...6L,2010A&A...514A..35N, 2013A&A...552A..20N}. First column: parameters 
of the hydrostatic initial model (luminosity 
$L_\star$, mass $M_\star$, effective temperature $T_\star$, metallicity $[$Fe/H$]$) together with the 
pulsation period $P$ of the piston at the inner boundary. $\Delta u_{\rm p}$, $f_{\rm L}$: velocity 
amplitude and luminosity-amplitude parameter of the inner boundary condition (piston) used to simulate the 
pulsating stellar interiors; $\Delta m_{\rm bol}$: resulting bolometric amplitude; $\kappa_{\rm 0}$: total 
dust opacity; $f_{\rm abs}$: fraction  assumed to be true absorption 
(see Sect.\,\ref{models}); $\langle\dot M\rangle$: mean mass-loss rate; 
$\langle u \rangle$: mean outflow velocity  at the outer boundary.}
\end{sidewaystable*}

\begin{figure}
 \resizebox{\hsize}{!}{\includegraphics{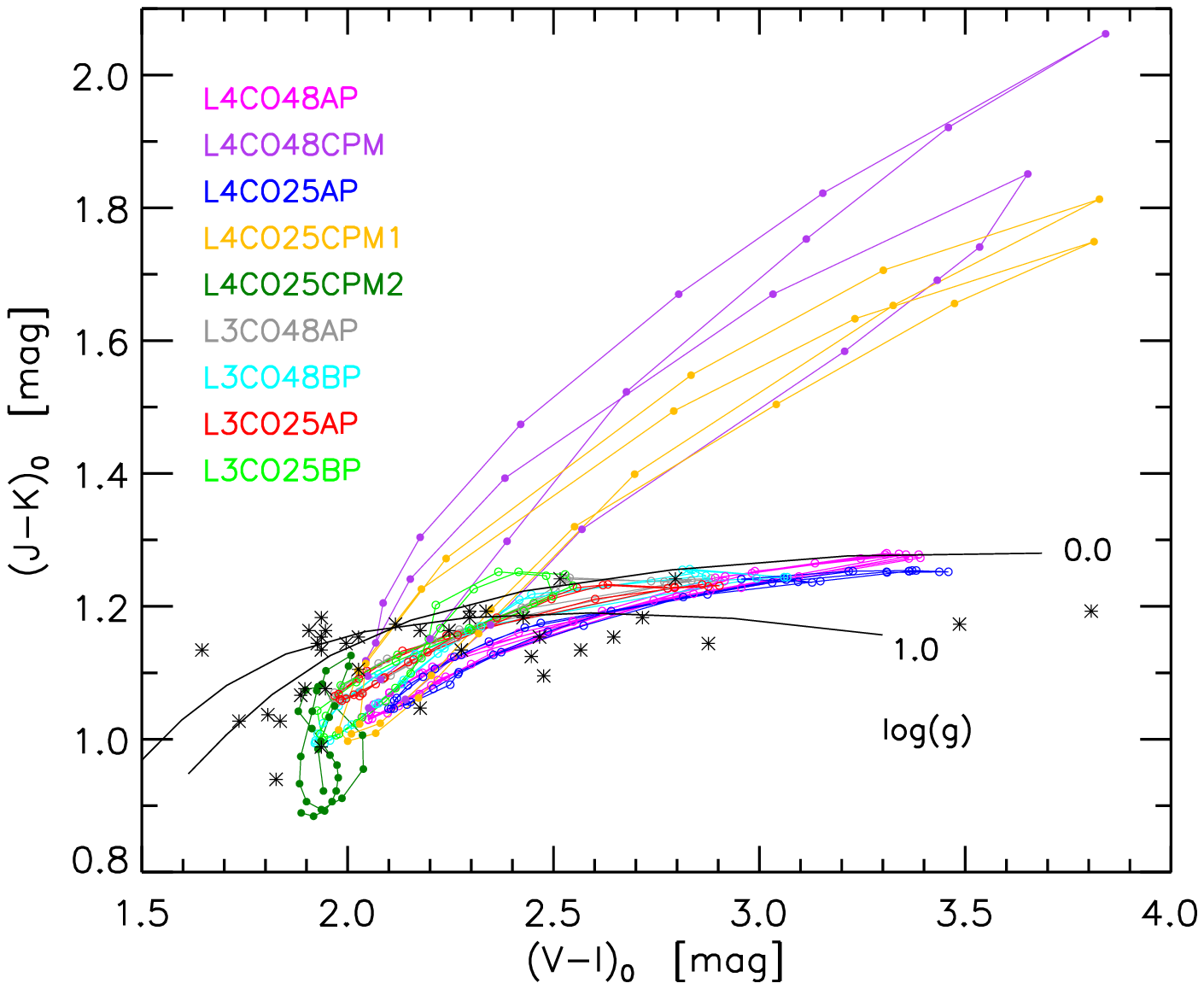}}
 \caption{Colour-colour diagram for the 47\,Tuc variables from \citet{2005A&A...441.1117L}, marked with asterisks. 
 Overplotted are temperature sequences of hydrostatic
 MARCS models for two different values of log\,$g$ (black solid lines), and the phase-dependent photometry for the dynamical model atmospheres.}
 \label{f:CCD}
\end{figure}

\begin{figure*}
 \resizebox{\hsize}{!}{\includegraphics[angle=90]{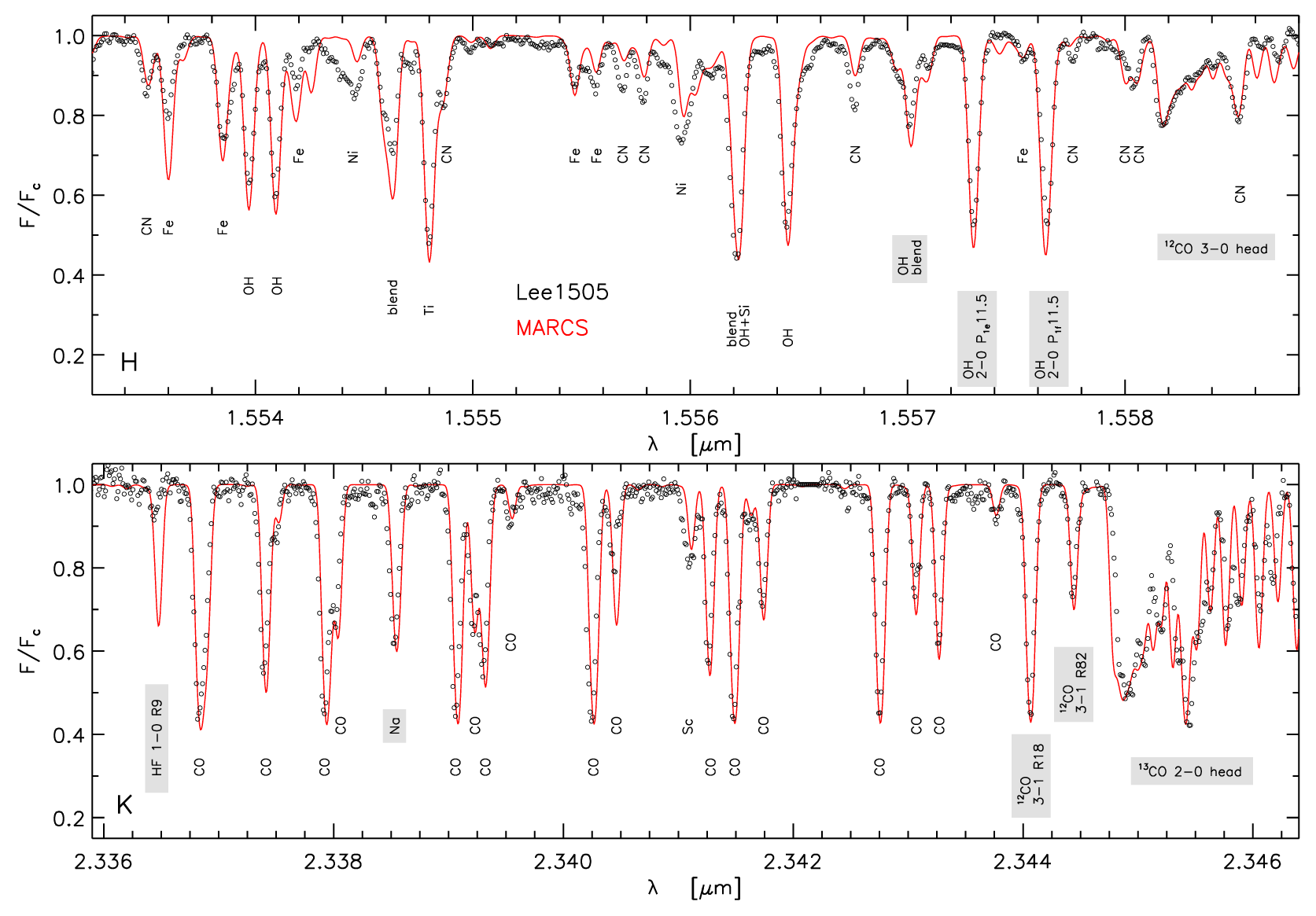}}
 \caption{Comparison of the observed spectrum of the non-variable 47\,Tuc star Lee 1505 with the best-fitting hydrostatic MARCS model (red lines). 
 The upper panel shows the $H$ band spectrum, the lower panel the $K$ band spectrum. Several prominent lines are identified. 
 Grey-shaded identifications mark those lines we used in our analysis (Table \ref{t:features}).}
 \label{f:specLee}
\end{figure*}

\begin{figure*}
 \resizebox{\hsize}{!}{\includegraphics[angle=90]{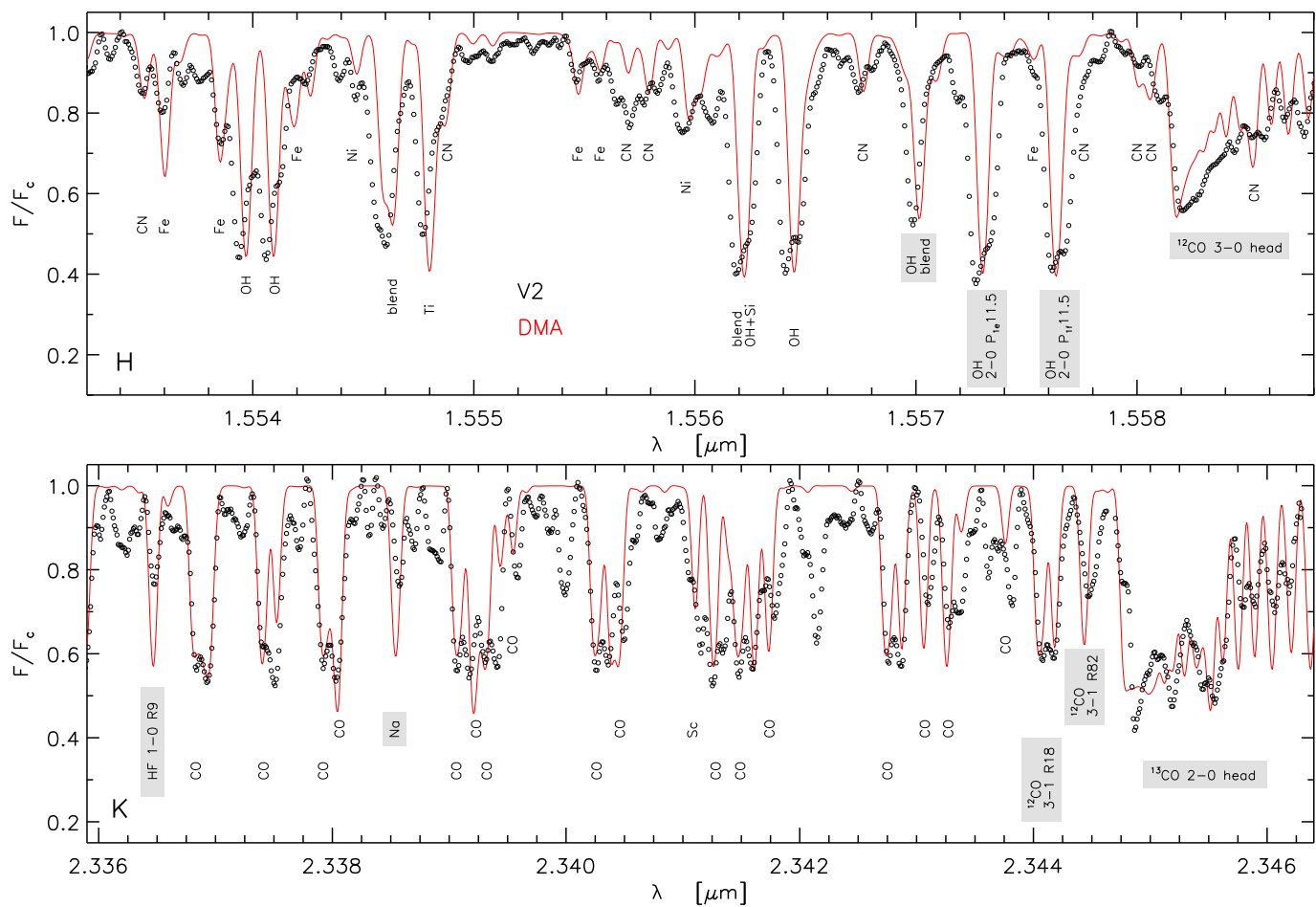}}
 \caption{Same as Fig.\,\ref{f:specLee} but for one of the 47\,Tuc variable stars, namely V2, compared with the best-fitting dynamical model L3C025BP at pulsation phase $\phi$\,=\,0.8.}
 \label{f:specV2}
\end{figure*}

\begin{figure}
 \resizebox{\hsize}{!}{\includegraphics[angle=90]{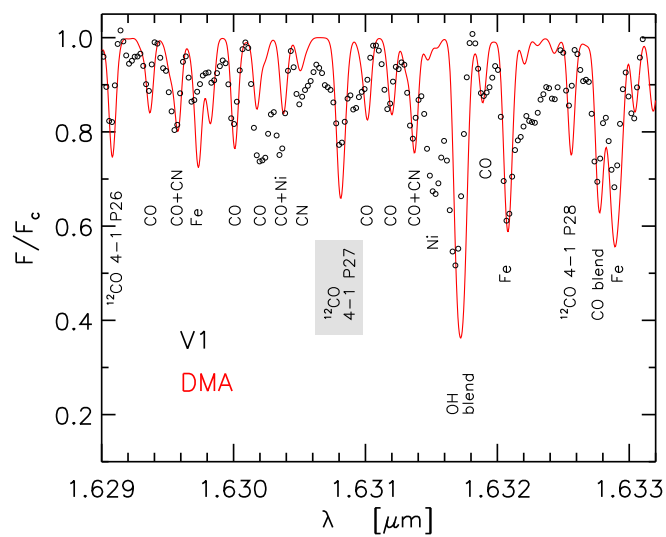}}
 \caption{Comparison of one observed spectrum of the 47\,Tuc variable V1, compared with the corresponding synthetic spectrum of model L4C025AP at minimum light during the light cycle (phase $\phi$\,=\,0.5). This additional wavelength range contains the 
$^{12}$CO 4--1 P27 line, which was used for investigating the EW variability in Sect.\,\ref{s:results2}. }
 \label{f:specvariability}
\end{figure}

\subsubsection{Constraints from photometry}
Before dealing with the high-resolution spectroscopic data, we
compared the synthetic broad-band photometry derived from our
hydrostatic and dynamical model atmospheres with the observational
data of the 47\,Tuc stars. This is done in Fig.\,\ref{f:CCD}. Along with
the variables from our sample, we also included 47\,Tuc cluster
giants taken from \citet{2005A&A...441.1117L}. For the variable
stars, mean values are given.  Unfortunately, 
$I_{C}$ photometry for the Lee stars listed in Table \ref{t:sample}
is not available, and 
they are not plotted in Fig.\,\ref{f:CCD}. However, we
can safely assume that the Lee stars will be found in the same range as the
other low-luminosity cluster giants.  The hydrostatic models for C/O\,=\,0.25 are represented
by two sequences for two different log\,$g$ values.
The complete light cycle
is shown for the dynamical models. 

The synthetic photometry of our models is in very good
agreement with the colour range covered by the 47\,Tuc stars except
for two models, L4C048CPM and L4C025CPM1, which strongly deviate
in $(J-K)$. 
Out of the three models in our mini-grid that develop winds, these two are 
the cases where the outflows are driven by radiation pressure due to true absorption by 
dust grains ($f_{\rm abs}$\,=\,1, see Sect.\,\ref{models}). The third wind model (L4C025CPM2) is, 
in contrast, based on the assumption that the driving force is due to pure scattering 
($f_{\rm abs}$\,=\,0, all other parameters identical to model L4C025CPM1), resulting in colours that 
are in good agreement with observations. This is consistent with the results of 
\citet{2013A&A...553A..20B},
indicating again that the circumstellar envelopes of O-rich AGB stars are rather transparent 
at NIR wavelengths, which makes radiation pressure by true absorption on dust grains an unlikely 
driving mechanism for these stars.\footnote{For solar metallicity, scattering of photons on Fe-free 
silicate grains seems to be a viable alternative as demonstrated by \citet{2008A&A...491L...1H} and 
\citet{2013A&A...553A..20B}. 
At the low metallicity and rather high effective temperatures of the 47 Tuc stars, however, this 
option is also unlikely, leaving the question about the driving mechanism open at present.} 
For illustrative
purposes, we decided to keep L4C025CPM1 in our analysis keeping in
mind its limitations in terms of the description of dust and its
non-fitting NIR colours.

We also calculated photometric amplitudes from our models. In the \mbox{$K$-band},
L4C025AP, L3C025AP, and L3C025BP all show a similar amplitude of
0.4--0.5 magnitudes.  L4C025CPM1 produces a $K$ amplitude of 0.9.
$V$ amplitudes of the 47 Tuc variables can be found in
\citet{2005A&A...441.1117L}, Table \ref{47tucsample}.  The $K$
amplitude is approximately 20\% of the visual amplitude
\citep[see][]{2005A&A...441.1117L}. Therefore, we expect total $K$
amplitudes around 0.6 to 0.9 magnitude for V1 to V3 and 0.4
magnitudes for V4. The calculated light amplitudes from our models
are thus in reasonable agreement with the observed values, but
only the L4C025CPM1 model produces enough amplitude to model
the light amplitude of the brightest stars in our sample, probably
due to variations in the dust absorption. However, as we saw in Fig.\,\ref{f:CCD},
this model results in an unrealistic variation in $(J-K)$.

\subsubsection{Analysis of the high-resolution spectra}
As a next step, we compared the observed and synthetic spectra for
the non-variable case (see Fig.\,\ref{f:specLee} for an illustrative
example). The best-fitting hydrostatic model was chosen on the basis
of the $(J-K)$ colour and the line strengths of the spectral features
listed in Table \ref{t:features} and also marked in
Fig.\,\ref{f:specLee}. For most spectral lines, both in the $H$-
and in the $K$-band region, the agreement between observed and
synthetic spectra is very good. In the $H$-band the largest
difference is found for the observed feature at the location of the
Ni line at 1.5544\,$\mu$m.  Quite obviously there is a blending
feature that has not been identified yet. Since it is quite broad,
we suspect that it is either a mix of several blending atomic lines
or a structure resulting from a molecular species. Besides that, we
see minor differences in some of the CN lines. This could indicate
either that the low N abundance is too low in our models or that 
there is inadequate line
data. In the $K$-band, no particular differences between observed
and modelled spectra were found except for a difference in the
strengths of the HF line. This is very likely due to an incorrect F
abundance in our models.  The F abundance of 47\,Tuc stars will be
discussed in a forthcoming paper (Hren et al., in preparation).
In any case, none of the deviations seen in the $H$-band has any effect
on the line strengths measurements in our study.

In Fig.\,\ref{f:specV2} we give an illustrative example for a
comparison between a variable star (V2) and a dynamical model. 
This spectrum was selected because it shows the
line doubling that \citet{2005A&A...432..207L} previously observed in V1, V2, and
V3.  In this case, the model
and the specific phase for comparison was selected by eye, with the
aim of reproducing the mentioned line doubling at least in the $K$-band.
However, V2 is in its post-maximum phase, while the model plotted
in Fig.\,\ref{f:specV2} is pre-maximum. The measured EWs
of the $^{12}$CO and $^{13}$CO bandheads differ by 0.2 and
0.1\,{\AA}, respectively.  The observed $(J-K)$ value is approximately
0.04 mag lower than the corresponding model value.  Still the
dynamical model fits the observation significantly better than any
of our hydrostatic models.

At this point, we excluded another model from our analysis,
namely L4C025CPM2. Although photometry derived from the model is in good
agreement with the colour range of the 47\,Tuc stars (Fig.\,\ref{f:CCD}),
the model spectra show strong emission lines over most of the
pulsation cycle. The emission lines are not visible in any of the observed spectra
and prohibit a meaningful measurement of line strengths
within our analysis scheme.

\subsection{Uncertainties of the EW measurements} \label{uncertain}

In the case of the EWs measured in the spectra of the
47\,Tuc stars we note three possible sources of uncertainty: 
the observational noise, the location of the continuum level, and
the effect of variations in blending features and differential
velocity fields in the stellar atmosphere. For the Gemini data, the
S/N of the spectra of the cluster stars (both variables and
non-variable stars) was between 150 and 250 except for the $H$ band
spectrum of V18, which had a S/N of 120. Due to the high quality
of the spectra and the low contamination by telluric lines, the
influence of the observational noise only plays a minor role
in the error budget.  This is a bit different for the Mount Stromlo
time series spectra that have a S/N around 50.

Defining a continuum level is quite a challenging task, in particular
for the coolest objects in our sample owing to not having enough
continuum points. In setting the level of the pseudo-continuum,
we followed the strategy described in Paper\,I. Observed and synthetic
spectra were handled the same way; i.e., we measured line strengths
in the synthetic spectra relative to a local pseudo continuum
and not the true continuum level to allow for a comparison with the
observations. We did some experiments by changing the continuum
level within an acceptable range among the highest points in the
spectra to estimate the error and found maximum values of
$\pm$0.02\,{\AA} in EW for individual lines and about
$\pm$0.04\,{\AA} for the band heads studied.

Even at high spectral resolution in the NIR, the analysis
of a spectral line without any blending from neighbouring features
is almost impossible. Again, the problem is more pronounced for the
cool variables at the tip of the giant branch. This confusion is
unavoidable, but since the effect is expected to be the same in the
observed and synthetic spectra, it should not be a major problem for
our analysis. A more significant challenge is
the variability of the line profiles within a pulsation cycle.
Problems here are twofold. First, the line centre moves in wavelength
according to the variable atmospheric velocity field.  As in
Paper\,I we handle this problem by shifting the spectra to laboratory
wavelength. A difficulty that arises for V1, V2, and V3 at some
phases is that the velocity shift is not necessarily the same for
all atomic or molecular transitions. This was taken into account
when positioning the lower and upper limits for measuring the
EWs. The second problem in this context are the changes
in the line width once secondary components appear. All the limits
for measuring the EW were set to the maximum width
possible without including the next feature of high or moderate
strengths. This selection was done on the basis of the hydrostatic
model spectra. However, in a few cases this width was not sufficient
to include all additional line components that emerge due to the
velocity field in the photosphere. An estimate of the size of this
uncertainty is not possible.

For the Mount Stromlo spectra, we occasionally have spectra of
the same star taken on two consecutive nights. Assuming that the
spectral variations on that time scale will be negligible, we used
a direct comparison of these two spectra for estimating the uncertainty
of our measured equivalent width. We consistently found uncertainties
of $\pm$0.03\,{\AA} for the results from this data set.

\section{Results}

\subsection{Comparing models with single-epoch observations}\label{s:results1}

The goals of our analysis are twofold.  One is to investigate
the spectroscopic differences between hydrostatic and dynamical
models based on a larger grid of models than in Paper\,I. As
outlined there, we find some basic characteristics of the spectral
changes with pulsation phase, in particular the loop of line EW
against $(J-K)$. The loops in line strength versus colour show that
the line-forming region has a different structure on the rising and
the descending branches. We could confirm this effect with our extended
set of dynamical model atmospheres.  The second goal is to compare
the modelling results with observational data to estimate several
basic stellar parameters for our sample of evolved red giants in
47\,Tuc. In the following, both goals will be pursued in parallel.
We use a fixed value for the
metallicity of the 47\,Tuc stars based on the considerable number
of measurements of this quantity available in the literature, and
we explore the sensitivity of the chosen spectral features
to fundamental stellar parameters. 

\subsubsection{Constraining the C/O ratios}
\begin{figure}
 \resizebox{\hsize}{!}{\includegraphics{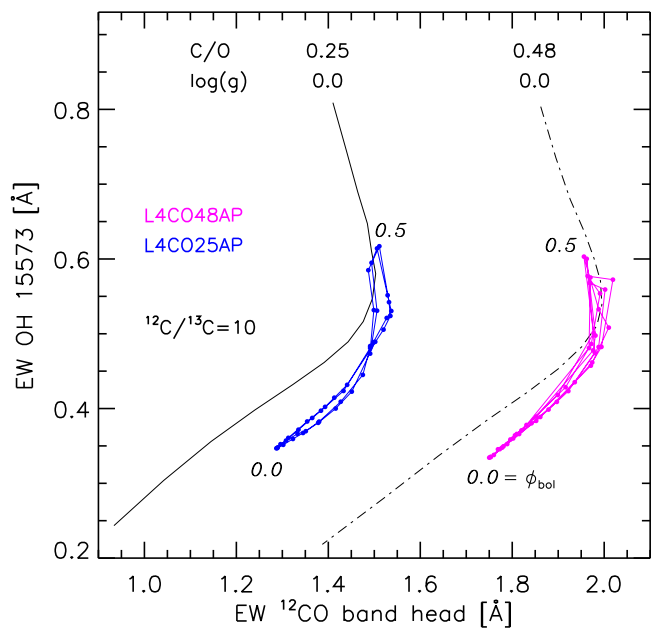}}
 \caption{Equivalent widths in [{\AA}] of the OH line at 15573\,{\AA} versus the strength of the $^{12}$CO band head. Black lines mark
 temperature sequences for log\,$g$\,=\,0.0 and two different C/O ratios. The two coloured lines show the behaviour of the dynamic models L4C048AP and
 L4C025AP. Maximum and minimum phases are indicated ($\Phi_{\rm bol}$\,=\,0.0 and 0.5, respectively). For all calculations a $^{12}$C/$^{13}$C ratio of 10 was assumed.}
 \label{f:C12OHmodel}
\end{figure}

In
Fig.\,\ref{f:C12OHmodel} we use two prominent features in the
$H$-band spectrum, namely an OH line and the $^{12}$CO 3-0 band
head, to study the dependence of the measured equivalent widths
on the C/O ratio. 
Plotting two feature strengths against each other
instead of plotting feature strength against $(J-K)$ has the
advantage of better separating the effects of temperature $T_{\rm eff}$ 
and log\,$g$. In this and all the following plots, the full
temperature range, 2600 to 4500\,K, is shown for the
hydrostatic models. 
For the dynamical models, we typically plot
two consecutive light cycles. 

For determining the C/O ratio we assume a scaled solar abundance
of oxygen. Since oxygen may be over- or under-abundant in the cluster 
stars relative to the scaled solar value, 
our measurement of the CO band head strengths nominally measures
only C*O. To explore this effect, we did some test calculations by changing the
oxygen abundance by +0.2 dex. The CO-band-head strength depends on
C/O, $T_{\rm eff}$, log\,$g$, and the oxygen abundance. 
Increasing either C/O or O/H leads in both cases to an increase in the feature
strengths. The OH lines,
in particular the OH blend near 1.557\,$\mu$m, are sensitive to the
same parameters\footnote{For the two strong OH
lines next to the CO band head, the effect of abundance changes on the
line strengths is fairly weak}. However, the dependency on
C/O is going the opposite way; i.e., the lines are getting weaker for 
a higher C/O ratio. Using  $T_{\rm eff}$ and log\,$g$ values from Table \ref{t:sample},
we can then constrain both C/O and O/H by fitting the OH lines and 
the CO band head at the same time. Since we are focussing on evolutionary abundance 
changes in this paper, we made no attempt to derive an accurate oxygen
abundance for these stars. However, we can clearly constrain the oxygen
abundance to be close to a scaled solar value since the model presented
in Fig.\,\ref{f:specLee} obviously fits both features nicely with the same
set of abundances. We estimate the
uncertainty to be less than 0.1 dex. Keeping this small uncertainty in mind
and assuming that the oxygen abundance is not changing along the giant branch
of 47 Tuc, we derive a C/O ratio for our sample stars in the following.

For a given log\,$g$ value we get a
very clear separation according to the C/O ratio of 0.25 and 0.48. 
It is also obvious that the dynamical models -- we
use here L4C025AP and L4C048AP as representative examples -- allow
a clear distinction between two different C/O ratios as well. The
log\,$g$ value of 0.0 chosen here for the MARCS models brings the
hydrostatic models closest to the dynamical ones. It is interesting to
note that the variability during the pulsation cycle in the dynamical
models produces effects that are rather similar to a sequence of hydrostatic
models with fixed log\,$g$ and different effective temperatures.  

\begin{figure}
 \resizebox{\hsize}{!}{\includegraphics{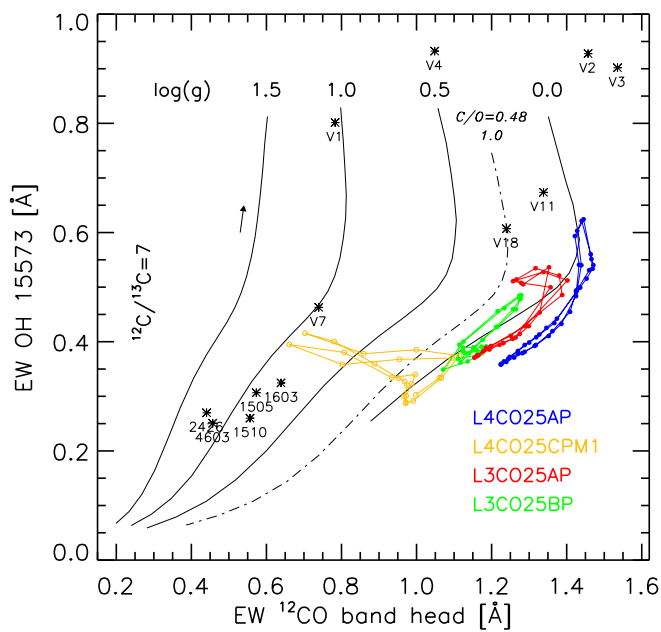}}
 \caption{Equivalent widths in [{\AA}] of the OH line at 15573\,{\AA} versus the strength of the $^{12}$CO band head. The 47\,Tuc stars (Table\,\ref{t:sample}) are marked
 with asterisks. Sequences of hydrostatic models (with varying $T_{\rm eff}$ but constant log\,$g$) are shown as black lines, the corresponding log\,$g$ values are labelled at the top of these lines. The small arrow
 indicates the trend of decreasing temperature. Coloured lines mark the location of selected dynamical models during the light cycles. For all
 models, the carbon isotopic ratio  $^{12}$C/$^{13}$C was set to 7. The C/O ratio is 0.25 if not stated otherwise.}
 \label{f:C12OH}
\end{figure}

Dynamical models with C/O$=$0.25 show similar or even
stronger $^{12}$CO band heads than hydrostatic models with
a log\,$g$ value of 1.0 as shown in Fig.\,\ref{f:C12OH}, where we
added the 47\,Tuc stars to the plot. Dynamical models with C/O$=$0.48
have been excluded from the plot since they are outside the range of the
observed EWs.  For illustrative purposes we have also included 
in Fig.\,\ref{f:C12OH} the range in EWs covered by the dynamical
model L4C025CPM1, which shows NIR colours that are far too red.

The non-variable 47\,Tuc stars all show similar values in Fig.\,\ref{f:C12OH}
and are located close to the temperature sequence of the
hydrostatic models with log\,$g$\,=\,1.0 and a C/O ratio of 0.25.
On the same sequence, we find the mildly variable star V7 and the long-period 
variable V1. However, an inspection of the spectrum of the
latter reveals very strong differences from a hydrostatic case,
therefore the agreement has to be seen as a coincidence. All
the other variables are found in the upper right-hand part of
Fig.\,\ref{f:C12OH}, clearly separated from the non-variable giants.

Figures \ref{f:C12OHmodel} and \ref{f:C12OH} strongly suggest that
the C/O ratio in the 47\,Tuc stars is below 0.48.  This is
also expected from stellar evolution theory considering their low
mass.  We therefore decided to focus on the models with C/O$=$0.25
in the following, namely L4C025AP, L3C025AP, and L3C025BP.

For completeness we mention that in all
cases a carbon isotopic ratio of 7 was used, but this parameter has
only a weak effect on the $^{12}$CO band-head strength.  The other
two OH features measured (see Table \ref{t:features}) show 
identical behaviour. 

\begin{figure*}
\sidecaption
\includegraphics[width=12cm,clip]{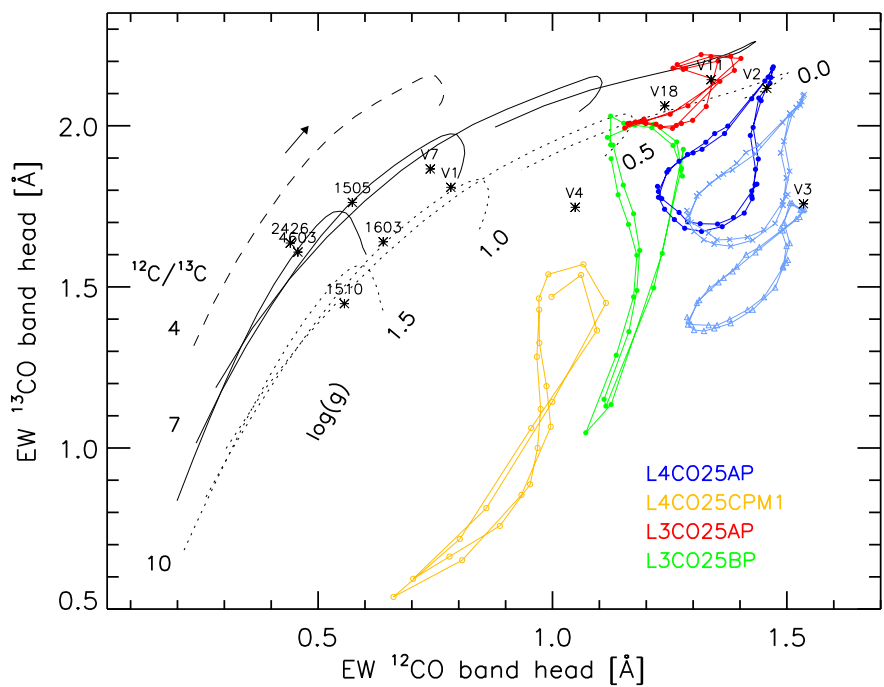}
\caption{Equivalent widths in [{\AA}] of the $^{13}$CO band head versus the $^{12}$CO band head. 47\,Tuc stars are marked with asterisks.
 Black lines indicate effective temperature sequences of hydrostatic MARCS models for various constant  log\,$g$ values and three different values for the
 carbon isotopic ratio $^{12}$C/$^{13}$C (as labelled). The temperature is decreasing in the direction of the small arrow. Coloured lines again show the location of the dynamical 
 models with a carbon isotopic ratio of 7 assumed. In the case of L4CO25AP, we show also the results for differing $^{12}$C/$^{13}$C ratios of 10 and 30 (light blue).}
 \label{f:C12C13}
\end{figure*}

\subsubsection{Constraining the $^{12}$C/$^{13}$C ratios}

In Fig.\,\ref{f:C12C13} we compare the strengths of the $^{12}$CO
and the $^{13}$CO band heads. Hydrostatic temperature sequences are
shown for three different $^{12}$C/$^{13}$C ratios and various
log\,$g$ values. The three dynamical models selected in the previous
step are plotted as well. As can be seen from Fig.\,\ref{f:C12C13},
the mass-losing model L4C025CPM1 does not reproduce the observed
strengths of the band heads either. To explore the effect of a
varying carbon isotopic ratio on the synthetic spectra of the
dynamical models, we calculated the corresponding values for
$^{12}$C/$^{13}$C ratios of 10 and 30 for model LC4C025AP, for all
the other dynamical models a value of 7 was used. The dynamical
models cover an area in this diagram that starts at somewhat
stronger $^{12}$CO band heads than the hydrostatic models for a
given depth of the $^{13}$CO band head and ends in the same area
as the low-temperature and low-surface gravity hydrostatic models.

All non-variable stars are found close to the hydrostatic temperature
sequences. As in Fig.\,\ref{f:C12OH} a log\,$g$ value close to
1.0 seems appropriate which agrees with the value estimated
from the colour, brightness, and mass of these stars (Sect.\,\ref{evaluation}).
For most stars, an isotopic ratio of 7 provides a good fit. Two
stars, Lee1510 and Lee1603, are fitted better by a slightly higher
ratio. Similar to our comparison of the EWs of the
OH line and the $^{12}$CO band head (Fig.\,\ref{f:C12OH}), V1 and
V7 are found near the hydrostatic sequences as well.  V2, V11, and
V18 are close to the low-temperature end of the hydrostatic
log\,$g\,=\,$0 sequence for $^{12}$C/$^{13}$C$\,=\,$10.  It is possible that
V4 could be brought into agreement with a hydrostatic
model with  $^{12}$C/$^{13}$C>10. We discuss this aspect below.
The uncertainty in the band strengths is typically
0.04\,{\AA}.  Therefore the measurement error has a very weak
effect on the results. Finally, V3 is found separated from the
other stars but within the range in EWs covered by
the dynamical models.

\subsubsection{Investigation of different line-forming regions}

\begin{figure}
 \resizebox{\hsize}{!}{\includegraphics{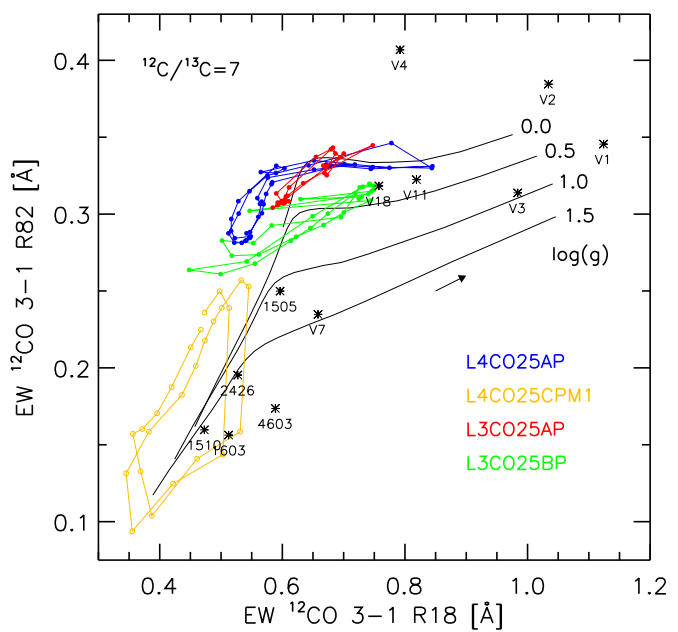}}
 \caption{Equivalent widths in [{\AA}] of the high excitation CO 3-1 R82 line versus the low excitation CO 3-1 R18 line. Same
 symbols and line styles as in Fig.\,\ref{f:C12OH}.}
 \label{f:R18R82}
\end{figure}

In Fig.\,\ref{f:R18R82} we compare two CO 3-1 lines of different
excitation energies. In the extended atmospheres of red
giants the R18 line and the R82 line will be formed at quite
different atmospheric depths. The excitation energy of R18 is
2794.0743\,cm$^{-1}$, while the excitation energy of R82 is 14826.3718\,cm$^{-1}$.
This is, in particular, relevant for
the variable stars where a study of the two lines permits  
comparison of atmospheric structure and
inherent velocity fields with models.  This approach has been applied to
AGB stars in various papers in the past
\citep[e.g.][]{1982ApJ...252..697H,1985PASP...97..994W,2005A&A...437..273N,2010A&A...514A..35N}.
All modelling results for this figure were calculated with a carbon
isotopic ratio of 7 and C/O\,=\,0.25.  However, neither parameter
has a significant effect on the results shown here.

Figure\,\ref{f:R18R82} is structured in a similar way to the previous ones.
The dynamical models are now closer to the hydrostatic relations,
suggesting that dynamical effects are not that prominent in these
lines. For effective temperatures above 3600\,K, the R82 line is
obviously a good indicator for $T_{\rm eff}$, while the dependency
on log\,$g$ is quite weak in this temperature regime. For lower
temperatures this parameter can be derived using the low excitation
R18 line. Lee 1505, Lee 1510, Lee 1603, and Lee 1505 behave as in
all the other diagrams. Lee 4603 is somewhat offset, but the reason for
this is not clear. The variables besides V7 are all found in the
upper right-hand part of the diagram. 
V4 has a remarkably strong R82 line.
Inspection of the line profile reveals a significant broadening due
to line doubling.

\begin{figure}
 \resizebox{\hsize}{!}{\includegraphics{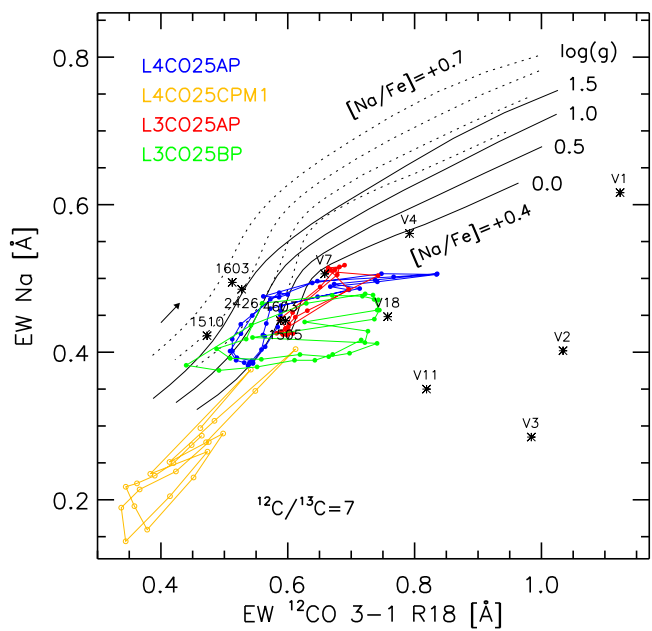}}
 \caption{Equivalent widths in [{\AA}] of the sodium line at 23383\,{\AA} versus the EW of the $^{12}$CO 3-1 R18 line. Same symbols and line styles
 as in Fig.\,\ref{f:C12OH}. Hydrostatic models are plotted for [Na/Fe]$=$+0.4 and +0.7, dynamical models for [Na/Fe]$=$+0.4, respectively.}
 \label{f:C13Na}
\end{figure}

\subsubsection{Sodium as an example for atomic lines}

The 2.3383 $\mu$m sodium (Na) line, located
within the spectral range covered by our $K$-band observation, 
was selected as representative of an atomic line.  Its
line profile has little contamination by other spectral lines (at
least in the hydrostatic case).  The EWs for this spectral line
derived from observations, hydrostatic and dynamical models are
plotted in Fig.\,\ref{f:C13Na} against the $^{12}$CO 3-1 R18 line
strengths. These two line strengths are related almost
linearly to each other making this result simpler than sodium line
strengths against, for instance, the $^{13}$CO band head strength. 
In either case, the derived conclusions are very similar.

Following published Na abundances we used an overabundance of Na
relative to Fe in our model calculations. \citet{2014ApJ...780...94C}
report [Na/Fe] values between 0.0 and 0.9 dex in 47\,Tuc stars,
with the majority of objects showing overabundances relative to iron between
0.2 and 0.9 dex.  This agrees with earlier results by other
authors \citep[e.g.][]{2013A&A...550A..34C}. Since the spectra of
the non-variable giants suggest a clear enhancement of the sodium
abundance in these stars, we decided to compute two series of
hydrostatic model spectra, one with [Na/Fe]$=$+0.4 and one with
[Na/Fe]$=$+0.7.  The findings from \citet{2014ApJ...780...94C}
suggest quite a large scatter in the sodium overabundance from
star to star, so that our usual assumption on the similarity of
abundances in our variable and non-variable sample stars is not
necessarily correct. The uncertainty in this case is also illustrated
by the difference observed between our non-variable giants: Lee
1510, Lee 1603, and Lee 2426 are nicely fitted with [Na/Fe]$=$+0.7
(for log\,$g$\,=\,1.0), while for the other Lee stars a value
slightly below [Na/Fe]$=$+0.4 seems to be more appropriate.

The variable stars V4 and V7 are found near the hydrostatic temperature
sequence for log\,$g$\,=\,0.0 and [Na/Fe]$=$+0.4. The other variables
all show a weaker sodium line, which would require [Na/Fe] values
around 0.0 (V18) or clearly subsolar (V1, V2, V3, V11). 
The results of \citet{2014ApJ...780...94C} make such low values
extremely unlikely. We therefore conclude that the Na line is
weakened in these variables by dynamical effects.  
We calculated synthetic spectra for the four
dynamical models also used in the previous comparisons with
[Na/Fe]$=$+0.4. The increase in EW for [Na/Fe]$=$+0.7 can be estimated
from the corresponding difference seen in the hydrostatic models.

An obvious difference in the behaviour of the dynamical models shown
in Fig.\,\ref{f:C13Na} compared with those shown in Fig.\,\ref{f:C12C13} is that the loops are
not as smooth in this diagram.  We think this is due to the appearance
of emission in the blue wing of the $^{12}$CO 3-1 R18
line at some phases. The effect is also visible in Fig.\,\ref{f:R18R82}.
The Na line does not show similar emission in the model spectra.
The models L3C025AP and L4C025AP, i.e. the ones with the smallest
piston amplitude, are found very close to the hydrostatic series
with log\,$g$\,=\,0.0 and [Na/Fe]$=$+0.4 throughout the whole
pulsation cycle.  Model L3C025BP with a slightly larger piston
amplitude exhibits a rather horizontal change in the diagram over
a large portion of the light cycle. In fact, the EW of the Na line
hardly changes between $\Phi=$0.9 and 0.4, while the $^{12}$CO 3-1
R18 line shows a sinusoidal variation with phase. As a result of
this difference, this dynamical model cannot be approximated by a
temperature sequence of hydrostatic models at constant surface
gravity. Finally, model L4C025CPM1 is clearly offset from the
hydrostatic ones towards low EW values. Weak line strengths, also
compared to the other dynamical models, are also found for the
$^{12}$CO 3-1 R82 line and the $^{12}$CO 3-1 R18 line as can be
seen in Fig.\,\ref{f:R18R82}. Emission components appearing in the
line profiles as mentioned above could be one cause but since the
Na line does not show obvious emission components in the dynamical
models, it seems to be more appropriate to describe this as a line
weakening, a well known phenomenon in LPVs
\citep[e.g.][]{1988ApJS...66...69C}.

\begin{figure*}
\sidecaption
\includegraphics[width=12cm,clip]{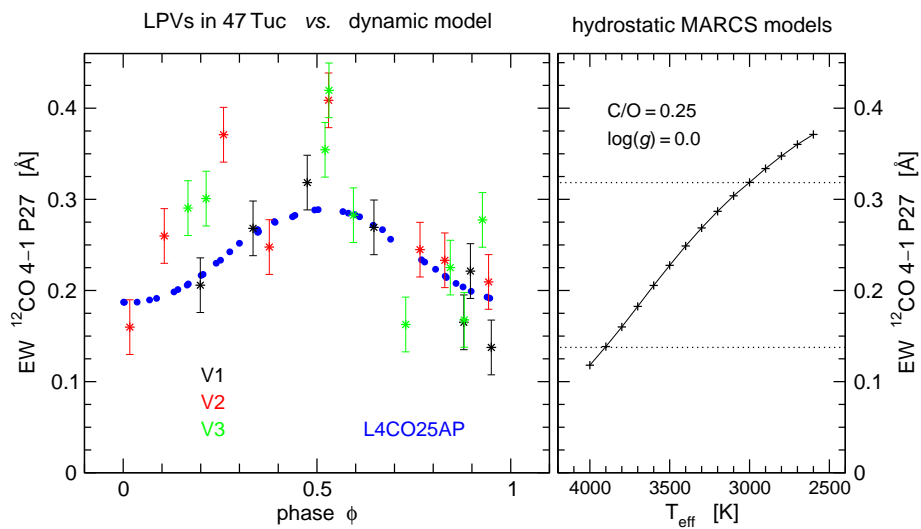}
\caption{\textit{Left:} Variation in the equivalent widths in [{\AA}] of the \mbox{$^{12}$CO\,4-1\,P27} line over the light cycle for the three 47\,Tuc LPVs with the largest
 photometric amplitudes (V1, V2, V3, colour coded). Also shown is the corresponding variation with pulsation phase as derived for the dynamical model L4C025AP. 
\textit{Right:} Line-strength variation
 of the same line for one $T_{\rm eff}$--sequence of hydrostatic models with log\,$g$\,=\,0.0.
The dotted lines mark the total range of equivalent widths measured for V1.}
 \label{f:timeseries}
\end{figure*}

\subsection{Variability of EW over a pulsation cycle}\label{s:results2}

The high S/N single-epoch observations used in the previous section
do not permit a direct evaluation of the variations with phase seen
in the model spectra.  We therefore made use of an additional time
series of spectra for our 47\,Tuc variables, which was obtained for
an earlier project \citep{2005A&A...432..207L}. Owing to the lower
resolution and the lower S/N, we decided to focus on a single line
that seemed to be least affected by neighbouring features, namely
the $^{12}$CO 4-1 P27 line (cf. Fig.\,\ref{f:specvariability}).
The strongest variations in EW were found in V1,
V2, and V3, the stars with the largest pulsation amplitudes. The
results are plotted against pulsation phase in the left-hand panel of
Fig.\,\ref{f:timeseries}. This plot combines data from
more than one light cycle, so that cycle-to-cycle variations have
to be considered as a source of additional scatter.  The other
47\,Tuc variables investigated -- V4, V11, and V18 -- show variations
with an amplitude of typically 0.1\,{\AA} with a similar dependency
on the pulsation phase, i.e. maximum line strength near light minimum
($\phi$\,$\approx$\,0.5).

The same trend can be observed for the dynamical models, most clearly
expressed in L4C025AP, which is also shown in Fig.\,\ref{f:timeseries}.
However, the amplitude of the dynamical model is somewhat less than
what is observed  in V1 to V3. Still, this variation corresponds to
a substantial change in temperature when compared with the
hydrostatic models. Such a comparison can be done with the help of
the right-hand panel of Fig.\,\ref{f:timeseries}.  This supports the
finding from Paper\,I that the dynamical model atmospheres can
qualitatively reproduce the observed variations in line strength.
Model L4C025CPM1, the model with the largest bolometric amplitude,
gives the same amplitude for the change in EW.

\section{Discussion}

\subsection{C/O and $^{12}$C/$^{13}$C ratios of red giants in 47\,Tuc} 

For the five non-variable cluster giants studied in this
paper (the Lee stars), the hydrostatic models provide fits to the observed
spectra sufficient to derive elemental abundances. As noted above,
the metallicity was set to a fixed value taken from the literature.

We start our analysis by constraining the C/O ratio using the
relations plotted in Fig.\,\ref{f:C12OH}. In the following we make the 
assumption that the five giants will have similar abundances; i.e., the
star-to-star scatter is small, although it is of course clearly
visible in the figures presented in the previous section. We also assume
that the log\,$g$ value derived from independent measurements is
adequate.  A value of C/O$\,=\,$0.25 seems to give a reasonably
good fit for the observations. Within the uncertainties of
the measurement, the calculated $T_{\rm eff}$ values of the Lee
stars (Table\,\ref{t:sample}) are consistent with the temperatures
of the best-fitting models in Fig.\,\ref{f:C12OH}.  Even allowing
an uncertainty of 0.5 in the value for the surface gravity, 
the error of the C/O ratio would be limited to 0.1. The measured subsolar C/O
ratio is consistent with a reduction of carbon on the surface by
the first dredge-up, as well as the oxygen overabundance and carbon
underabundance in pre-RGB stars in 47\,Tuc measured by
\citet{2005A&A...433..597C}.

Next we determine the $^{12}$C/$^{13}$C ratio with the help of the
quantities plotted in Fig.\,\ref{f:C12C13}. The combination of the
$^{12}$CO and $^{13}$CO
band heads has the convenient property that a change in C/O
shifts the relation between the two feature strengths along the
path for a constant isotopic ratio; i.e., it has no effect on the
determination of the latter quantity. The carbon isotopic ratio can
therefore be determined with an accuracy of $\pm$1 (for the range
given here). For Lee 1505, Lee 2426, and Lee 4603, we find a ratio
of 7, and for Lee 1510 and Lee 1603 a ratio of 10. For dwarfs in this
cluster, \citet{2005A&A...433..597C} find isotopic ratios clearly
above 10, while their sample of subgiants shows $^{12}$C/$^{13}$C
values between 9 and 12. Attributing this reduction to first dredge
up suggests that the main sequence value was lower than the
solar isotopic ratio, a consequence of primordial pollution with
CN burning products in the cluster.  The values for the subgiants
are only slightly higher than the values we found for our RGB stars.
This is consistent with the finding of \citet{2003ApJ...585L..45S}
that the isotopic ratio of carbon stays nearly constant above the
RGB bump. However, the mild reduction we see is likely due to
extra-mixing on the RGB \citep[e.g.][and references
therein]{2003ApJ...582.1036N,2008ApJ...677..581E,2009A&A...502..913L}.

\subsection{Abundance determination in LPVs} 

The low stellar
masses of the 47\,Tuc AGB stars mean we can safely assume that the
surface abundance is not altered by third dredge up
\citep[e.g.][]{2003PASA...20..389S}. Combined with the
small scatter of the abundances found for our five 
red giants, we expect to find abundances similar to the red giants 
for the AGB stars.
This allows us to study the effect of the dynamics on the
abundance determination.

We compare different observed spectroscopic features with results
of hydrostatic models, identifying discrepancies and/or inconsistencies.
The fundamental stellar parameters, i.e. mass, luminosity, and
$T_{\rm eff}$, derived for the LPVs in the 47 Tuc sample
suggest log\,$g$ values between $-$0.2 and 0.5 as
appropriate. For each spectroscopic feature we explore
how the feature strength compares with the prediction from the model.  

We 
start with the OH 15573 line versus the $^{12}$CO
band head, used for determining C/O 
(Fig.\,\ref{f:C12OH}). V1 and V7 are found close to a
temperature sequence of hydrostatic models
with log\,$g$\,=\,1.0; i.e., the $^{12}$CO band
head comes out too weak or the OH line is too strong to place them
on a sequence corresponding to an appropriate log\,$g$. Surface
gravities derived from mass, colour, and luminosity, as done here,
therefore indicate a C/O ratio of significantly less than 0.25 in
both cases (to shift a temperature sequence with a given log\,$g$ to the left
in Fig.\,\ref{f:C12OH}). This is, however, unrealistic, indicating that the lines
are affected by dynamics. In this context it is interesting to recall that the
dynamical model L4CO25CPM1 predicts a change over the pulsation
cycle reaching from the area of hydrostatic models with log\,$g\,=\,$0
to those with log\,$g\,=\,$1.0. In this sense, a difference between
the observed EW and the expected log\,$g$ value might
be reproducible by dynamical models.

V11 and V18 are located within the expected log\,$g$ range, so a
hydrostatic model would lead to a consistent C/O ratio of 0.25.  
For V2, V3, and V4 we measure very strong OH lines, most likely due to
line doubling as outlined below. All three stars are found beyond the
end of our model sequences. 
From the stellar parameters derived in Sect.\,\ref{47tucsample}, the log\,$g$
value of V4 is expected to be closer to 0, so the star is
somewhat offset to the left of the corresponding hydrostatic sequence.
The CO band-head strength of V2 and V3 is consistent with
the estimated surface gravities around log\,$g$\,=\,$-$0.2 and a C/O ratio
of 0.25 for these two stars. A C/O ratio of 0.48 (indicated in Fig.\,\ref{f:C12OH}
by the dash-dotted line for  log\,$g$\,=\,1.0) is only possible if the surface
gravity of the two variables is significantly higher than expected. However,
since the concept of determining stellar parameters for long-period variables
has to be seen with caution, the C/O ratio determined in this way comes
with some uncertainty. Nevertheless, we note the consistency of the findings
with the assumption of a rather constant C/O for all objects on the upper giant
branch of 47\,Tuc.

It is interesting to compare this result with the pulsation phases
at the time of observation. Since the light curves of V7, V11, and
V18 are not regular enough to allow for determining a reliable
pulsation phase, we restrict ourselves to V1, V2, V3, and V4.
We found that V2 and V3 show line strength that is very
comparable to a hydrostatic model of similar log\,$g$. Their pulsation
phases are 0.35 and 0.18, respectively.  V1 and V4, which both show
an offset, were observed at phases 0.9 and 0.8, respectively. The
dynamical model L4C025CPM1 shows a similar offset at some
phases. However, the model reaches the maximum offset near light
minimum, i.e. phase 0.5.  Up to this point we have focussed on the log\,$g$ value in our comparison.
However, we also have to take a closer look at the temperature,
even though the concept of defining a $T_{\rm eff}$ value is 
questionable for the highly extended atmospheres of AGB variables.
Still we immediately note that stars like V1, V2, V3, or V4 are
found close to the low-temperature end of our hydrostatic sequences,
corresponding to temperatures of 2600\,K or even less.  Observed NIR
colours, however, suggest a significantly higher temperature at the
time of the observations (see Table \ref{t:sample}). The difference
of 900\,K is also far more than what would be expected at any time
during the light cycle. Furthermore, no spectra of the four stars 
were obtained near minimum light where
we would expect the star to be reddest.  Therefore, while the
log\,$g$ values seem to be consistent with the observed
EWs, the temperature values we estimate from
NIR photometry are not.  The reason for this can be easily
seen in the upper panel of Fig.\,\ref{f:specV2}. The OH line at
15573\,{\AA} we study here is clearly broadened in V2 by a second
component; i.e., we see a case of beginning line doubling. This leads
to an increased EW of the line.

In summary, for this feature we find that
C/O ratios derived from the two-feature-strength plot presented in
Fig.\,\ref{f:C12OH} under the assumption of a given log\,$g$ value
are either as expected or lower. Our observations suggest a
dependency of this offset on phase.  The feature strengths are
inconsistent with an effective temperature derived under hydrostatic
assumptions.

In the case of the comparison between the $^{12}$CO and $^{13}$CO
band head (Fig.\,\ref{f:C12C13}), V1 and V7 show similar behaviour
to the previous two features studied; i.e., the measured EWs
correspond to a log\,$g$ value significantly higher than
expected. We note the bending of the hydrostatic temperature sequences
at the low-temperature end. V1 is indeed at the end of the log\,$g=$1.0
sequence for an isotopic ratio $^{12}$C/$^{13}$C$=$7, not on the
$^{12}$C/$^{13}$C$=$10 sequence. Therefore, this is completely consistent
with the findings from Fig.\,\ref{f:C12OH} regarding the log\,$g$
value. The same is true for V4, which is again found somewhat above
the low-temperature end of the log\,$g=$0.5, $^{12}$C/$^{13}$C$=$7
sequence.  V2, V11, and V18 give band strengths similar to the
predictions of a hydrostatic model with log\,$g=$0.0. The only
variable with a different behaviour is V3, which probably has too
weak a $^{13}$CO band head. The most trivial explanation that the
star has a higher carbon isotopic ratio would require either a
dredge up of $^{12}$C on the AGB, which is very unlikely in these
stars and which would also manifest itself in a different C/O ratio,
or the absence of extra-mixing on the first giant branch, which poses
the question why this single star has undergone a different evolution
or had a different primordial isotopic ratio. While possible, we
find also a deviating behaviour of V3 in another spectral feature
as discussed below. For the comparison of the two band heads we end
up with the same and consistent conclusion as for the comparison
of the $^{12}$CO band head and the OH line. Again, an effective
temperature consistent with the measured EWs is
inconsistent with the colours of these stars.

\subsection{Atmospheric structure and dynamics of LPVs}
Comparison of the EWs of the two $^{12}$CO 3-1 lines
we plotted in Fig.\,\ref{f:R18R82} provides insight into the stellar
atmospheric structure since both lines are affected by abundance changes in a
very similar way. In particular, the differential velocity fields
in the stellar atmosphere resulting from the pulsation significantly
change the line profile of these moderate excitation CO lines as
illustrated by the line doubling visible in the lower panel of
Fig.\,\ref{f:specV2} \citep[cf.][]{1982ApJ...252..697H}. Inspection
of these two lines in the $K$-band spectra of our seven variables reveals
a variety of line profiles which we summarize in Table \ref{t:profiles}.

\begin{table}
\caption{Qualitative description of the line shapes of the $^{12}$CO 3-1 lines R18 and R82.}
\label{t:profiles}
\centering
\begin{tabular}{l l l}
\hline\hline
Object-ID & R18 & R82 \bigstrut[t] \bigstrut[b]\\
\hline
V1 & doubled, red weak & slightly broadened \bigstrut[t]\\
V2 & doubled, similar strength & broadened\\
V3 & doubled, blue weak & slightly broadened\\
V4 & single, broadened & broadened\\
V7 & single & single\\
V11 & single, & slightly broadened\\
 & slightly broadened & \\
V18 & single & single \bigstrut[b]\\
\hline
\end{tabular}
\end{table}

We first discuss the stars showing neither line doubling
nor strong line broadening, i.e. V7, V11, and V18. V7, a star behaving
very consistently in all other features discussed, shows a value
that places it close to the hydrostatic log\,$g\,=\,$1.5 sequence;
i.e., for our hydrostatic approach either its R82 line is too weak
or both lines are too weak. Considering its typical location along
the temperature sequence in the other diagrams, the latter case is
more likely. This finding is very interesting since it indicates
that even stars that are thought to be quasi-hydrostatic owing to
their small pulsation amplitude still can exhibit clear signatures
of a non-hydrostatic atmospheric structure. In this context we also have
to note the slight broadening of these lines in V11 compared
to a hydrostatic case. However, the line strengths in the spectrum
of V11, and also in V18, are not outstanding and are consistent
with the results from the other features where these stars show
no strong deviation from the hydrostatic models.

In the spectral features discussed above the occurrence of line
doubling led to a location in our two-feature diagrams consistent
with our derived log\,$g$ value but inconsistent with the colour
temperature. In the case of the two CO lines, this is true for V2,
but not for V3 and V4, both deviating strongly from the hydrostatic
log\,$g\,=$\,0 sequence. At this point, the limitations of fitting
dynamical stars with hydrostatic models are clearly illustrated.
However, none of our dynamical
models studied here reaches the line strengths observed in the large amplitude
47\,Tuc variables. For completeness we note that V1, which shows
indications of line doubling in both CO lines, behaves consistently
with what we found for the other spectral features.

Since the primordial scatter of the Na abundance is quite large in
47\,Tuc, the Na line in our spectra is only of limited use in 
estimating the dynamical effects on the abundance determination.
However, the strong offset of V1, V2, and V3 from the other giants
in Fig.\,\ref{f:C13Na}
suggests that the Na line is weaker in the dynamical case than in
hydrostatic models of similar stellar parameters. The behaviour of
the dynamical models suggest that there might be phases where the
sodium line strengths approach the hydrostatic case, but to check
this spectroscopic time series covering a whole light cycle for at
least one of the large amplitude variables would be needed.

\subsection{Behaviour of the dynamical models}
While the limited set of dynamical models used in this study led to some
qualitative improvements compared to hydrostatic models, it did not provide
satisfactory quantitative fits for all investigated features of 
the long-period variables of 47\,Tuc. However, the four
different sets of parameters for the models analysed in
detail in Sect.\,\ref{s:results1} permit a study of the effects of these
parameters on the EWs of the synthetic spectral
features.  We noted above that for some spectral features
temporal variations of the dynamical models closely resemble a hydrostatic
temperature sequence at constant log\,g, while in other features
they strongly deviate from the hydrostatic models. The deviation from
the hydrostatic case is more obvious for the band heads
studied -- see in particular Figure \ref{f:C12C13} -- than for
individual atomic or molecular lines.  The spectra of
V3 and V4 support the possibility that such deviations in the
$^{13}$CO band head strength from the hydrostatic case
occur in real LPVs.

A main result of Paper\,I was the discovery of an asymmetry in the variation
of the feature strengths between the rising and the descending
branch of the light curve. As a result, the models -- and also
observed miras -- produce a loop in a diagram showing feature strength vs.
colour. 
In the present paper we show
that these loops are found also in diagrams relating two feature
strengths with one another. 
The loops are reminiscent of loops in colour-colour diagrams due to
phase-shifted variations of molecular features originating in different
layers \citep{2013A&A...553A..20B}.
Although one can recognize some differences in the location and extension
of such loops our limited set of models does not permit the identification
of systematic dependencies with any model parameter.

In dynamical models abundance differences not only lead to an overall
shift of the curves (as found for the hydrostatic models) 
describing EW change over a
pulsation cycle but can also affect the shape of that loop.
This is clearly seen for model L4C025AP in Fig.\,\ref{f:C12C13}
for curves of carbon isotopic ratios 7 (dark blue), 10, and
30 (light blue). The shape of the loop changes for an isotopic
ratio of 30 producing a
double loop structure. Since 
all these three calculations are based on the very same atmospheric
structure,
this demonstrates the strong effect of shifting the line forming region in geometrical 
distance depending on the adopted isotopic ratio in the case of dynamical atmospheres.
A similar change is seen for the C/O ratio in Fig.\,\ref{f:C12OH}.

\section{Conclusions}

In this paper we derived abundances for a small sample of red, 
non-variable giants in the globular cluster 47\,Tuc.  Assuming no
abundance changes due to third dredge up or other mixing processes
in the cluster's low mass AGB stars we tested various options for
deriving elemental abundances for long period variables. Our 
high-resolution spectra of the AGB variables in 47\,Tuc reveal that
the objects with the largest variability amplitudes show
effects on the line profiles due to atmospheric velocity fields
that cannot be reproduced by hydrostatic models. As illustrated in
Fig.\,\ref{f:specV2}, dynamical model atmospheres can in principle
produce similar line components \citep[see also][]{2005A&A...437..273N}.
However, none of the nine configurations of the dynamical models tested here
(Table \ref{t:dmaparameters}) gave a good fit to the entire
observed high-resolution spectra.  Nevertheless, the dynamical model
fits the observed spectrum of the large-amplitude variable V2 much better than any
hydrostatic model, thereby indicating the potential of dynamical model 
analysis of high-resolution LPV spectra.

In our analysis, we concentrated on comparing observed and
modeled equivalent widths of selected spectral features. While the
range of values covered by hydrostatic and dynamical models shows
some overlap with the observed values, neither hydrostatic nor
dynamical models were able to provide a consistent fit for any of
the four most variable stars, V1, V2, V3, and V4. Some success for
a fit with a single hydrostatic model could be achieved for the
mildly pulsating stars V11 and V18, although -- as for the other
variables -- the corresponding temperature of a hydrostatic model
with the same equivalent widths and the same log\,$g$ value is much
lower than what would be expected from the stars' colours.  The
fitting of V18 by a single hydrostatic model is remarkable since
the star shows significant dust mass loss, and therefore one would
expect an atmospheric structure that shows clear signs of stellar
pulsation\footnote{We note that the dust signatures of this star
could be the result of a higher mass loss rate in the past, so the
current mass loss rate could be lower. See the discussion on the
evolutionary status of V18 in \citet{2006ApJ...653L.145L}.}.  For
the third small amplitude LPV in our sample, V7, a model producing
good fits was not found.

We can compare our findings with the work by \citet{2007MNRAS.378.1089M},
who made a similar attempt to derive element abundances from
NIR spectra of some LMC AGB stars. They identify the time
around visual light minimum as the best one for this task, while
spectra obtained close to light maximum could not be fitted with
synthetic spectra. In the intermediate mass stars studied in their
paper, the shock enters the atmosphere around phase 0.7 to 0.8 and
leaves it again near phase 0.4.  The spectroscopic time series of
the 47 Tuc variables V1 to V3 presented in \citet{2005A&A...432..207L}
suggest the presence of line doubling, hence of a shock in the
atmosphere, between phases 0.75 and 0.2, i.e.~very
similar to the intermediate mass stars.  Unfortunately, we do not
have any Phoenix observations obtained near light minimum, therefore
we cannot directly test the finding of \citet{2007MNRAS.378.1089M}
that the minimum is the best phase to measure element abundances.
Our coverage of a complete pulsation cycle in the $^{12}$CO 4-1 P27
line (Figure \ref{f:timeseries}) does not favour a particular pulsation
phase in terms of a better model fit. The range of measured equivalent
widths of that line in the 47\,Tuc variables agrees with the range
covered by the hydrostatic models for log\,$g$\,=\,0, although it again
requires that quite low temperatures are reached by the variables,
temperatures too low to have a counterpart in the near-infrared colours.

The dynamical models tested here did not properly reproduce 
the equivalent widths for the 47\,Tuc variables. One
difficulty lies in the selection of the parameters of the
starting model based on observational quantities. We showed
that agreement in luminosity, colour indices, and metallicity, and an
approximate agreement in the $K$ amplitude is not sufficient to
obtain a model that satisfactorily fits the strengths of observed spectral 
features. It is not clear whether this indicates a problem in the
model structure or the need to improve the model selection by
including further stellar parameters. An obvious problem 
is a realistic description of the mass loss in these stars.
Some of the 47\,Tuc stars clearly show a dusty mass loss, 
but none of the three dynamical models that develop winds produces realistic
spectra. The two models where the outflows are driven by true absorption on
dust grains do not reproduce the observed near-infrared colours.  The remaining
model, if assuming pure scattering on dust grains as a driving force, shows 
problematic emission features.
It is reasonable to assume that also under the conditions in the
47 Tuc AGB stars dust formation in the stellar atmosphere will
affect the atmospheric structure and extension. This, again, would
influence the line strengths in the spectra. To tackle this problem,
a better understanding of mass loss and dust formation in these 
low-luminosity and low-metallicity AGB stars is clearly necessary.

Keeping in mind the variety we see in the observations and 
the uncertainties in the dynamic models regarding the fundamental 
parameters and the wind mechanism, 
it is clear that the study presented in this paper and
in Paper I is a limited first effort on the highly complex topic
of determining element abundances and isotopic ratios for 
large-amplitude variable stars. Hydrostatic models of low log\,$g$ and
an effective temperature lower than what is consistent with photometry
can be useful for extracting the stellar composition from the
high-resolution spectra of AGB variables, but the results come with
considerable uncertainties that are far beyond observational errors.
In general, selecting stars with smaller light amplitudes eases the
fitting process.  However, our study also shows that even mildly
pulsating stars can have spectral line strengths that cannot
consistently be fit with hydrostatic models.

Abundance offsets between values derived from hydrostatic models 
and the corresponding variable star observations need to be validated by
further investigations. The path followed here -- using AGB variables
for which the abundances can be predicted from other,
non-variable sources -- seems to be a good choice for this validation
process. But also for the next step, deriving abundances directly from AGB
variables, ensembles of such objects in stellar clusters could be the
preferred target since the expected similarity of the abundance
pattern can be used to average out the dynamical effects
acting in individual stars.

\begin{acknowledgements}
This work has been supported by the Austrian Science Fund (P23737-N16,
P21988-N16) and the Swedish Research Council.  The National Optical Astronomy Observatory is operated
by the Association of Universities for Research in Astronomy, Inc.
under cooperative agreement with the National Science Foundation.
This work is based in part on observations obtained at the Gemini
Observatory, which is operated by the Association of Universities
for Research in Astronomy, Inc., under a cooperative agreement with
the NSF on behalf of the Gemini partnership: the National Science
Foundation (United States), the National Research Council (Canada),
CONICYT (Chile), the Australian Research Council (Australia),
Minist\'{e}rio da Ci\^{e}ncia, Tecnologia e Inova\c{c}\~{a}o (Brazil),
and Ministerio de Ciencia, Tecnolog\'{i}a e Innovaci\'{o}n Productiva
(Argentina).
The authors wish to thank Dick Joyce for doing most of the data
reduction for the Mount Stromlo spectra.  Oscar Straniero provided
fruitful discussion on the atmospheric  mixing.  This publication
used data products from the Two Micron All Sky Survey, which
is a joint project of the University of Massachusetts and the
Infrared Processing and Analysis Center/California Institute of
Technology, funded by the National Aeronautics and Space Administration
and the National Science Foundation.  This research also made use
of the SIMBAD database, operated at the CDS in Strasbourg, France, and
NASA's Astrophysics Data System Bibliographic Services.

\end{acknowledgements}

\bibliographystyle{aa} 
\bibliography{dynabun} 

\end{document}